\documentclass{aastex63}
\accepted{May 5, 2022}

\submitjournal{ApJ}

\shorttitle{Breakthrough Listen broadband signal searches}
\shortauthors{Gajjar et al.}
\usepackage{graphics,graphicx,url,verbatim,xspace}
\usepackage{comment}
\usepackage[caption=false]{subfig} 
\graphicspath{{./}{figures/}}

\begin{document}

\title{{Searching for broadband pulsed beacons from 1883 stars using neural networks}}
\newcommand{\UCB}{Breakthrough Listen, University of California Berkeley, Berkeley, CA 94720, USA}
\newcommand{\UCBastro}{Department of Astronomy, University of California Berkeley, Berkeley, CA 94720, USA}
\newcommand{\SSL}{Space Sciences Laboratory, University of California Berkeley, Berkeley, CA 94720, USA}
\newcommand{\SWIN}{Centre for Astrophysics \& Supercomputing, Swinburne University of Technology, Hawthorn, Australia}
\newcommand{\GBTadd}{Green Bank Observatory,  West Virginia, 24944, USA}
\newcommand{\OXF}{Astronomy Department, University of Oxford, Keble Rd, Oxford, OX13RH, United Kingdom}
\newcommand{\NIJ}{Department of Astrophysics/IMAPP,Radboud University, Nijmegen, Netherlands}
\newcommand{\ATNF}{Australia Telescope National Facility, CSIRO, PO Box 76, Epping, Australia}
\newcommand{\HOU}{Hellenic Open University, School of Science \& Technology, Parodos Aristotelous, Perivola Patron, Greece}
\newcommand{\USQ}{University of Southern Queensland, Toowoomba, QLD 4350, Australia}
\newcommand{\SETI}{SETI Institute, Mountain View, CA 94043, USA}
\newcommand{\KZA}{University of Malta, Institute of Space Sciences and Astronomy, Msida, MSD2080, Malta}
\newcommand{\PWJD}{The Breakthrough Initiatives, NASA Research Park, Bld. 18, Moffett Field, CA, 94035, USA}
\newcommand{\BPF}{The Breakthrough Initiatives, NASA Research Park, Bld. 18, Moffett Field, CA, 94035, USA}
\newcommand{\cornell}{Cornell Center for Astrophysics and Planetary Science and Department of Astronomy, Cornell University, Ithaca, NY 14853, USA}
\newcommand{\PENN}{Department of Astronomy and Astrophysics, Pennsylvania State University, University Park PA 16802}
\newcommand{\MIT}{Massachusetts Institute of Technology, Cambridge, MA 02139, USA}
\newcommand{\NAU}{School of Informatics, Computing, and Cyber Systems, Northern Arizona University, Flagstaff, AZ, 86011}
\newcommand{\Curtin}{International Centre for Radio Astronomy Research, Curtin Institute of Radio Astronomy, Curtin University, Perth, WA 6845, Australia}
\newcommand{\FOOTBALLERS}{Department of Physics and Astronomy, University of Manchester, UK}

\correspondingauthor{Vishal Gajjar}
\email{vishalg@berkeley.edu}

\author[0000-0002-8604-106X]{Vishal Gajjar}
\affiliation{\UCB}
\author{Dominic LeDuc}
\affiliation{\UCB}
\author{Jiani Chen}
\affiliation{\UCB}
\author[0000-0003-2828-7720]{Andrew P. V. Siemion}
\affiliation{\UCB}
\affiliation{\SETI}
\affiliation{\FOOTBALLERS}
\affiliation{\KZA}
\author[0000-0001-7057-4999]{Sofia Z. Sheikh}
\affiliation{\UCB}
\affiliation{\SETI}
\author[0000-0002-7461-107X]{Bryan Brzycki}
\affiliation{\UCBastro}
\author[0000-0003-4823-129X]{Steve Croft}
\affiliation{\UCB}
\affiliation{\SETI}
\author[0000-0002-8071-6011]{Daniel Czech}
\affiliation{\UCB}

\author[0000-0003-3197-2294]{David DeBoer}
\affiliation{\UCB}

\author{Julia DeMarines}
\affiliation{\UCB}

\author{Jamie Drew}
\affiliation{\BPF}

\author[0000-0002-0531-1073]{Howard Isaacson}
\affiliation{\UCB}
\affiliation{\USQ}

\author[0000-0003-1515-4857]{Brian C. Lacki}
\affiliation{\UCB}

\author{Matt Lebofsky}
\affiliation{\UCB}

\author{David H.\ E.\ MacMahon}
\affiliation{\UCB}

\author[0000-0002-3616-5160]{Cherry Ng}
\affiliation{\UCB}
\affiliation{\SETI}
\affiliation{Dunlap Institute for Astronomy \& Astrophysics, University of Toronto, 50 St.~George Street, Toronto, ON M5S 3H4, Canada}

\author{Imke de Pater}
\affiliation{\UCBastro}

\author[0000-0002-6341-4548]{Karen I. Perez}
\affiliation{Department of Astronomy, Columbia University, 550 West 120th Street, New York, NY 10027, USA}

\author[0000-0003-2783-1608]{Danny C.\ Price}
\affiliation{\UCB}
\affiliation{\Curtin}

\author[0000-0002-5389-7806]{Akshay Suresh}
\affiliation{\cornell}

\author{Claire Webb}
\affiliation{\MIT}

\author{S. Pete Worden}
\affiliation{\BPF}

\begin{abstract}
The search for extraterrestrial intelligence at radio frequencies has largely been focused on continuous-wave narrowband signals. We demonstrate that broadband pulsed beacons are energetically efficient compared to narrowband beacons over longer operational timescales. Here, we report the first extensive survey searching for such broadband pulsed beacons towards 1883 stars {as a part of the Breakthrough Listen's search for advanced intelligent life}. We conducted 233 hours of deep observations across 4 to 8 GHz using the Robert C. Byrd Green Bank Telescope and searched for three different classes of signals with artificial (or negative) dispersion. We report a detailed search --- leveraging a convolutional neural network classifier on high-performance GPUs --- deployed for the very first time in a large-scale search for signals from extraterrestrial intelligence. 
{Due to the absence of any signal-of-interest from our survey}, we place a constraint on the existence of broadband pulsed beacons in our solar neighborhood: $\lesssim$1 in 1000 stars have transmitter power-densities $\gtrsim$10$^5$ W/Hz repeating $\leq$500 seconds at these frequencies. 
\end{abstract}


\section{Introduction} 
\label{sec:intro}
\subsection{Searching for broadband signals from ETI}
\label{sect:broadband_justification}
The search for extraterrestrial intelligence (SETI) is one of the most profound ventures to understand humanity's place in the cosmos. \cite{Tarter:2003p266} argues that electromagnetic waves from technology built by extraterrestrial intelligences (ETIs), especially at radio frequencies, still serve as one of the best possible ways to detect evidence of extraterrestrial life via their ``technosignatures''. {Following suggestions from \cite{1959Natur.184..844C}}, a major fraction of SETI efforts have been focused on locating extremely narrowband signals across a limited fraction of the radio spectrum \citep{Drake:1961bv,Verschuur1973,Tarter:1980p1516,1986Icar...65..152V,Horowitz:1993p1523,2013ApJ...767...94S,Harp_2016,Tingay_2016,Tingay_2018,Enriquez:2017,Harp:2018apj,Pinchuk_2019,price2020,Sheikh_2020,Gajjar_21_BLGCI}.

The time--frequency formulation of the uncertainty principle suggests that ETI signals (intentional or unintentional) could also occupy the corner of parameter space corresponding to temporally-limited broadband signals (i.e., transients, see \citealt{CoEk79}). \cite{Clancy_1980_wideOverNarrow} was the first to discuss the advantage of a broadband beacon rather than a CW narrowband beacon, from the perspective of the ETI transmitter. Frank Drake, in \cite{SETI_pioneers}, stated that ``{\itshape The most rational ET signal would be a series of pulses that would be evidence of intelligent design.}'' Project Cyclops \citep{NASA:2003p185}, one of the most ambitious and detailed design studies for technosignature searches, also suggested broadband pulses as one class of likely ETI beacons. 

The primary limitation of broadband signals is their susceptibility to propagation effects such as dispersion, scattering, and scintillation due to the intervening interstellar medium (ISM)  \citep{Shostak_1995_wide_bandwidth_SETI,Blair_2010_effect_of_ISM_on_beacon}. However, by studying astrophysical broadband pulsed emitters such as pulsars \citep{Hewish:1968}, rotating radio transients \citep{mclaughlin:2006} and fast radio bursts (FRBs; \citealt{Lorimer:2007p5652}), we can characterize these effects from the ISM and actually use them as a feature to identify true ETI beacons. For example, dispersion effects from the ISM cause broadband signals from astrophysical sources to exhibit a predictable early arrival at higher frequencies compared to lower frequencies. \cite{Demorest_2004_negative_DM} and \cite{Siemion:2010p6845} suggested that ETI might send broadband pulsed signals with \textit{negative} dispersion as a means of adding artificiality: broadband signals which appear to arrive earlier at lower frequencies compared to higher frequencies, contrary to natural phenomena\footnote{This strategy scores perfectly on the Ambiguity axis of the 9 Axes of Merit for Technosignature Searches \citep{sheikh2020nine}, as there are no known natural confounders.}. 

Despite these advantages, to the best of our knowledge, no detailed targeted searches have been performed for such signals, likely due to limitations in compute resources. Distributed computing is one potential solution: \texttt{ASTROPULSE} \citep{vonKorff:2010p3275} conducted blind searches for 0.4 $\mu$sec long pulses with the help of thousands of volunteers to overcome these limitations. In the last decade, radio SETI has entered a new era with the advancement in computing power enabled by Graphics Processing Units (GPUs) and widely available machine learning (ML) algorithms. The Breakthrough Listen (BL) Initiative is a US \$100 million 10-year project to conduct the most sensitive, comprehensive, and intensive search for technosignatures on other worlds \citep{ism+17,wds+17,Gajjar_19_astro2020}. The BL program aims to utilize these advances in compute power and algorithms to explore a range of possible ETI beacon types which have never been investigated before. 

\subsection{AI in the search for ETI}
\label{sect:AI_to_look_for_ETI}
Radio SETI searches must contend with large data volumes and a complex background of radio frequency interference (RFI) of anthropogenic origin, which could make it difficult to identify a real signal from ETI. A typical radio SETI survey defines a target signal type, and then aims to search for that signal class by designing a ``filter'' -- for example, \texttt{turboSETI} looks for CW narrowband drifting signals of artificial origin \citep{Enriquez:2017}. Such filter-based searches often produce millions of hits --- for example, \cite{Enriquez:2017} and \cite{price2020} reported around 29 and 51.7 million initial hits from their filter-based searches towards 692 and 1138 stars, respectively. These signals almost entirely originate from human technology such as mobile phones, wireless communication technologies, and satellites. This prevalence of anthropogenic interference has tightly constrained the filters that can be employed, and consequently, the ETI signal morphologies that can be investigated with this method.  

Advances in deep neural networks (DNNs) and in particular, convolutional neural networks (CNNs) for image classification, can help assess the validity of these large numbers of hits and reduce the quantity to a level more suited for manual inspection. \cite{Brzycki_2020_ML_SETI}  designed a CNN classifier that can be trained to identify relatively weak injected ETI signals ``drowning'' in strong RFI using synthetic datasets. Similarly, \cite{Harp:2018apj} demonstrated the usefulness of various neural network classifiers in identifying seven different kinds of likely ETI beacons. 
Recently, \cite{Pinchuk_margot_2021} also demonstrated the usefulness of CNN-based classifiers for discriminating RFI from true ETI signals. \cite{zhang18} used a hybrid approach where a CNN-based classifier identified FRB candidates, and then a filter-based dispersion check reduced the number of false positives. Here, similar to \cite{zhang18}, we have used a hybrid approach. We have developed a state-of-the-art GPU-accelerated pulse detection pipeline named \texttt{SPANDAK} (discussed in detail in Section \ref{sect:SPANDAK}) to carry out quick filter-based searches across a large number of observations. We have also developed and deployed a CNN-based classifier to eliminate a large fraction of false positives, presented in detail in Section \ref{sect:ML_pipeline} of the Appendix. With the help of the CNN classifier, we were able to reduce the large number of false positives generated from the \texttt{SPANDAK} pipeline by over $97\%$. 

Here, we report the first-ever targeted search for broadband pulsed ETI beacons towards 1883 stars. In Section \ref{sect:intro_aDM}, we introduce three types of artificially dispersed broadband signals, and we compare their power budget with CW narrowband signals in Section \ref{sect:power_budget}. Details of our observations are presented in Section \ref{sect:observations}. We report our results in Section \ref{sect:results} with the implication of our findings discussed in Section \ref{sect:discussion}, and Section \ref{sect:conclusion} lists our final conclusions. 

\section{Artificially dispersed signals}
\label{sect:intro_aDM}
\begin{figure}
    \centering
    \includegraphics[scale=0.5]{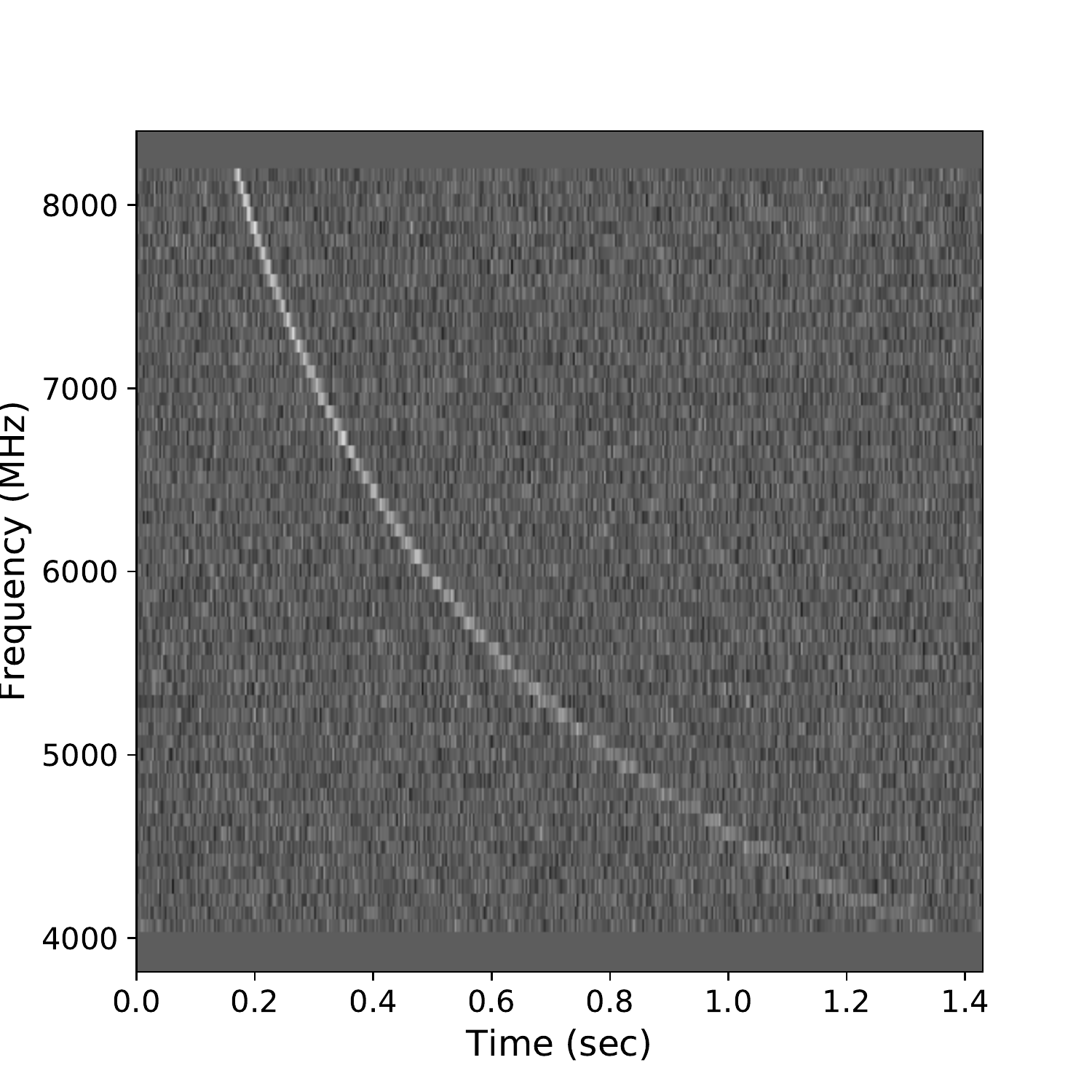}
    \caption{A simulated positively dispersed (pDM) signal, with a dispersion measure of 6000 pc-cm$^{-3}$, showing an example of natural dispersion across our observed frequencies. The signal is stored as a two-dimensional array [$\mathcal{V}$, T] of 64$\times$4096 bins.}
    \label{fig:pDM_example}
\end{figure}

Broadband signals passing through the interstellar medium experience dispersion due to cold ionized plasma. As mentioned in Section \ref{sec:intro}, the frequency-dependent refractive index of this plasma will cause signals at higher radio frequencies to arrive earlier than signals at lower frequencies (see Figure \ref{fig:pDM_example}). A typical broadband signal can be presented in a two-dimensional array of frequency against time -- [$\mathcal{V}_N$, $T$] -- with $N$ frequency channels and $T$ time samples. The dispersion delay for a broadband signal can be expressed for this two-dimensional array as, 
\begin{equation}
    t^H_i ~ = ~ \kappa \times DM \times \Big(\frac{1}{\nu_i^2} - \frac{1}{\nu_H^2} \Big) ; ~ i \in [0,\mathcal{V}_N] ~\& ~ t_i \in [0,T]. 
\end{equation}
Here, $\kappa$ is a constant, DM is the dispersion measure, $t^H_i$ is the arrival time bin of the signal in channel $i$ -- which is at a frequency of $\nu_i$\,Hz -- compared to the highest reference frequency of $\nu_H$\,Hz. Figure \ref{fig:pDM_example} shows an example of such a naturally-dispersed broadband signal represented in a two-dimensional array of 64$\times$4096 bins. We will refer to these naturally occurring signals as positively dispersed (pDM) signals.  

\begin{figure}[h]
    \centering
    \includegraphics[scale=0.6]{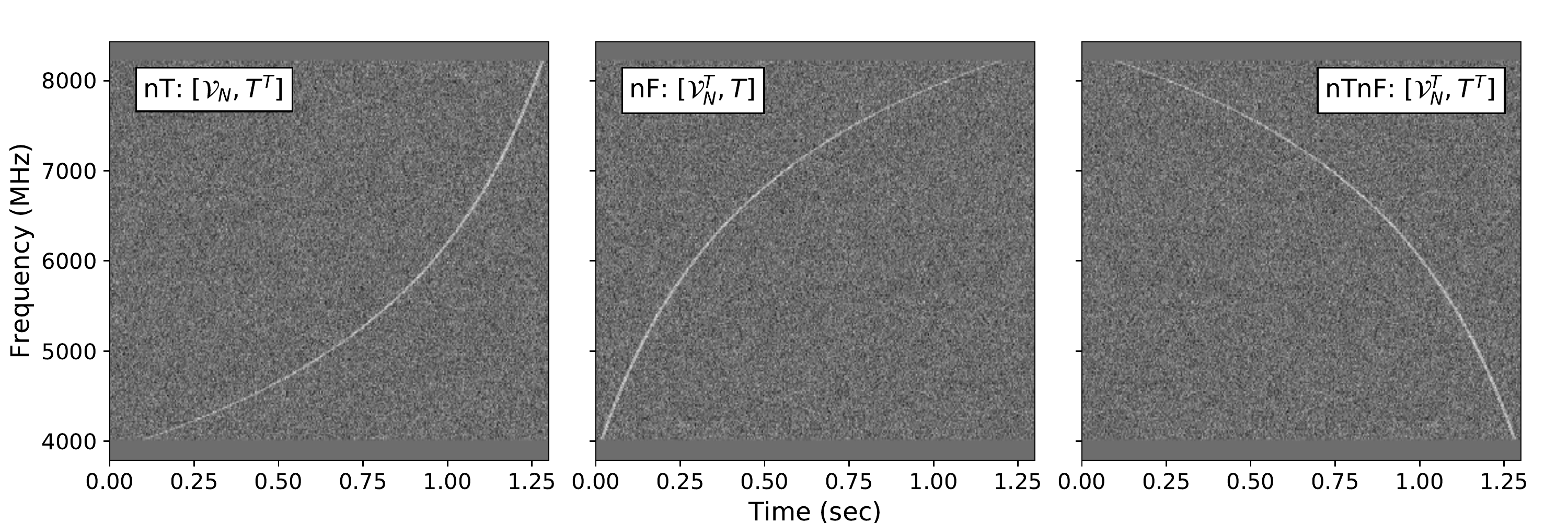}
    \caption{Simulated examples of the three different kinds of artificially-dispersed transients (aDM) included in this project. We performed a search for signals from these three classes with the \texttt{SPANDAK} pipeline across 4 -- 8\,GHz from data collected with the Breakthrough Listen program towards 1883 stars. These signals are negative time aDMs, negative frequency aDMs, and negative time and frequency aDMs from left to right, respectively.}
    \label{fig:negDM}
\end{figure}

It is possible that ETIs would be aware of the existence of astrophysical pDM signals, but choose to use pDM signals as beacons by adding obviously artificial features \citep{Demorest_2004_negative_DM,Siemion:2010p6845}. One method of modifying pDM signals would be  the use of extremely short pulses: the \texttt{ASTROPULSE} project searched for pulses of the order of 1$\mu$sec  \citep{korpela_SETIatHOME_2009, vonKorff:2010p3275}. However, as \cite{vonKorff:2010p3275} has suggested, other astrophysical sources are also known to emit such short pulses; FRBs have recently been shown to have sub-microsecond structures \citep{Nimmo_2021_M81_microstruct,majid_2021_M81_microstruct}. Another form of artificiality could be added by producing pDM signals exhibiting a repeating non-physical sequence, for example, a Fibonacci series or fundamental frequency of any well-known element \citep{Sullivan_91_ETI_pulse_periods}. 

Here, we postulate instead three simple variants of the pDM signal which are not yet known to occur in nature (see \citealt{Gajjar_21_BLGCI} for detail). As stated previously, broadband signals can be represented as two-dimensional arrays (pDM $\in$ [$\mathcal{V}$, T]). Transposing either or both of the axes of these arrays adds artificiality to the pDM signals. By transposing the time axis, one can artificially produce a signal which arrives at lower frequencies first and then gradually drifts to higher frequencies. Similarly, the frequency axis can be transposed such that the shape of the pDM signal appears to be reversed. Thus, there are three different classes of axis-transposed signals that one can search for: negative time (nT: [$\mathcal{V}$,T$^T$]), negative frequency (nF: [$\mathcal{V}^T$,T]), and both negative time and frequency (nTnF: [$\mathcal{V}^T$,T$^T$]). We will refer to these signals as artificially-dispersed signals (aDM), shown in Figure \ref{fig:negDM}. 

\section{Power budget of a broadband pulsed beacon}
\label{sect:power_budget}
As mentioned in Section \ref{sec:intro}, almost all previous SETI surveys have searched for CW narrowband signals. In this section we compare the total energy spent on a transmitter broadcasting a) a CW narrowband beacon against b) a broadband transient aDM beacon, as we described in Section \ref{sect:intro_aDM}. The output power of any transmitter, also known as Effective Isotropic Radiated Power (EIRP$_{out}$; \citealt{Enriquez:2017}), can be expressed as
{\begin{equation}
    {\rm EIRP_{out}~=~P_{a}\;G_{ET} ~[Watts]}
    \label{eq:pout}
\end{equation}
Here, ${\rm P_{a}}$ is the power provided to an antenna in Watts and ${\rm G_{ET}}$ is the gain of the ET antenna. For transmission occurring across a bandwidth ${\nu}_{\rm{ET}}$, we define power spectral density (PSD) in units of Watts/Hz. 
\begin{equation}
{\rm PSD_{ET} =  \frac{P_{a}}{\nu_{\rm{ET}}} ~ {[Watts/Hz]}}    
\end{equation}}
The transmitting antenna can be of any form; a single giant dish or multiple antennas spread across a large area operating as a phased-array. As an illustration, let us consider that the transmitting antenna is similar to GBT with antenna gain ${\rm G_{ET}~=~{4\pi{A^{eff}_{ET}}}/{\lambda^2}}$, where A${\rm^{eff}_{ET}}$ is the effective aperture and $\lambda$ is the wavelength. For a GBT-sized telescope operating at 6\,GHz with an aperture efficiency of 70\%, we expect a gain of $\sim$3$\times$10$^7$. To set a fiducial power, we will assume that our example ETI aims to send signals which can be detected at a distance D of 1000\,pc by a receiver with similar gain to the transmitter antenna. The minimum required power density for an ETI transmission (${\rm PSD_{ET},min}$) to be detected depends on its directionality and other characteristics of the signal; however, we shall assume a perfect alignment of transmitter and receiver for simplicity. For a broadband signal with bandwidth similar to the receiver bandwidth (i.e., ${\rm \nu_{ET}\geq{\nu_r} }$), 
this can be expressed as 
\begin{equation}
{\rm PSD_{ET,min,broad} ~ = ~ \Big(\frac{S}{N}\Big)_{min} \frac{SEFD_r}{G_{ET}} \frac{4\pi{D^2}}{\sqrt{\nu_{r}{\Delta\tau}}} ~ \approx 2\times10^5 \, \Bigg(\frac{D}{1000\,pc}\Bigg)^2~ [Watts/Hz]}     
\label{eq:P_et_broad}
\end{equation}
Here, ${\rm (S/N)_{min}}$ is the minimum required signal-to-noise ratio for detection (assumed to be 10), SEFD$_r$ is the system-equivalent-flux-density of the receiver, which is 10\,Jy at 6\,GHz for GBT with receiver bandwidth $\nu_{r}$ of 4\,GHz, ${\rm \Delta\tau}$ is the temporal width of the broadband signal (assumed to be 0.3\,ms). For CW narrowband signals, we expect ETIs to concentrate all the output power into a narrow frequency, ideally ${\nu}_{\rm ET}\leq$1\,Hz. For such signals, the minimum detectable power density can be given as
\begin{equation}
{\rm PSD_{ET,min,narrow} ~ = ~ \Big(\frac{S}{N}\Big)_{min} \frac{SEFD_r}{G_{ET}\nu_{ET}} \sqrt{\frac{\Delta\nu_r}{\tau_{obs}}} 4\pi{D^2} ~ \approx 2\times10^7\,\Bigg(\frac{D}{1000\,pc}\Bigg)^2 ~ [Watts/Hz] }
\end{equation}
Here we have assumed a channel bandwidth $\Delta\nu_r$ similar to the transmitter bandwidth of 1\,Hz, and we have set $\tau_{obs}$, the length of observation, to be 5 minutes for our receiver. For this example, we find that the power density required to send a detectable 0.3 millisecond broadband beacon is lower than the power density required for a detectable continuous narrowband signal lasting 5 minutes. 

The other disadvantage of sending a narrowband beacon is that it requires the sender to choose a transmission frequency. This limitation does not exist for broadband beacons, as their signals are likely to exist across several GHz, increasing the signal's chance of detection. However, one of the best \textit{advantages} of sending a narrowband signal is the ability for the receiver to integrate the incoming signal, which allows beacons with significantly lower power levels to be received. 
For example, integrating for 1 hour allows us to detect power densities on the order of 6$\times$10$^6$\ Watts/Hz. Furthermore, it is likely that an ETI might send a signal with $\nu_{ET}\leq$1\ Hz which would increase their peak power density requirement by several orders of magnitude. Similarly, an ETI might send a ``comb'' of narrowband signals separated by a few MHz, which would increase their chance of detection, as suggested by \cite{Shostak_1995_wide_bandwidth_SETI}.  

\begin{figure}
    \centering
    \includegraphics[scale=0.6]{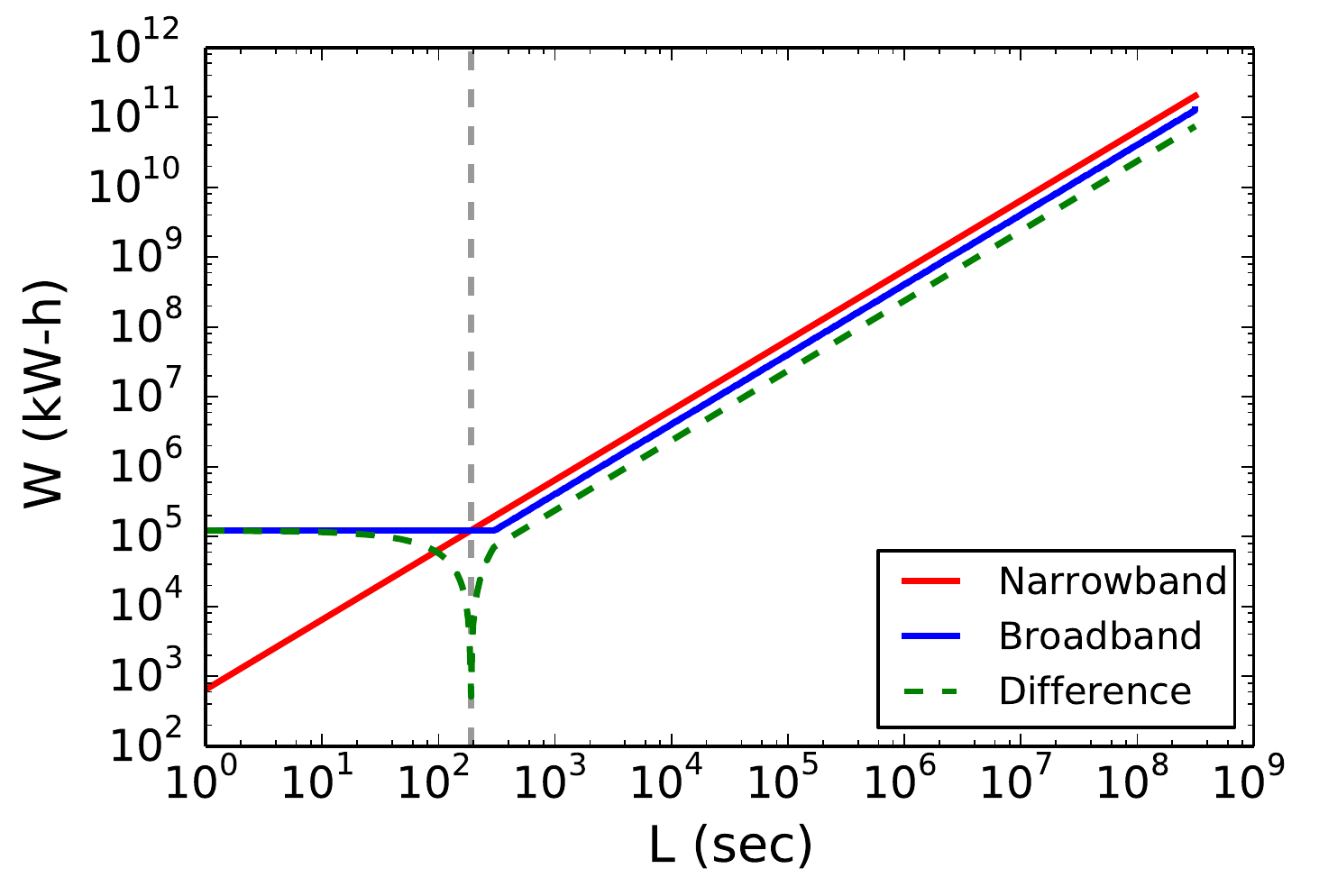}
    \caption{Power budget comparison for two types of ETI beacons. For the narrowband beacon (red), we have assumed a ``comb'' of narrowband signals separated by 40\,MHz. For the broadband beacon (blue), we have assumed a periodicity of 300\,seconds. The vertical dashed line shows the transition point where the total power consumed for the narrowband beacon equals that for the broadband beacon, which for our assumed example occurs around 200 seconds. The absolute difference between the broadband pulsed beacon and the narrowband beacon is also shown explicitly with the green dashed line. It is apparent that the longer a transmitter operates (past the transition point) the better the broadband pulsed strategy fares versus the CW narrowband strategy.}
    \label{fig:power_budget}
\end{figure}

We can calculate the total minimum required power budget (or total power consumed) and compare them between these two classes of signals for operating such transmitters as
\begin{equation}
  {\rm  W_{min}~=~PSD_{ET,min}\nu_{ET}L ~ ~ [kW \cdot h]}
\end{equation}
Here, $L$ is the operating time of the transmitter. Figure \ref{fig:power_budget} shows that on shorter operating timescales, narrowband transmitters have an advantage; however, over longer operating timescales, the narrowband transmitter power budget will likely be similar or higher than that of a broadband transmitter. The Drake Equation \citep{shklovskii_sagan_book} implies that the chances of success in a SETI search will largely depend on the lifetime of the transmitter. In other words, longer-lasting transmissions have a better chance of being detected. Figure \ref{fig:power_budget} also shows the power budget difference between the two methods, indicating that the power cost of operating a narrowband transmitter versus a broadband pulsed transmitter progressively widens with transmitter operating time. Thus, we speculate that for any sufficiently-advanced ETI wishing to transmit over a timescale of several hundred years, sending broadband signals is more desirable than sending CW narrowband signals. {\cite{2010AsBio..10..475B} carried out a detailed estimates of the capital and operation costs for pulsed ETI beacon transmitters. They concluded that short ($\mu$sec) pulses repeating around 1000 times a second serve as the most cost effective transmission strategy.}  


Recently, \cite{Gajjar_21_BLGCI} conducted one of the most comprehensive blind surveys towards the Galactic Center. Along with CW narrowband signals, \cite{Gajjar_21_BLGCI} also searched for broadband artificially dispersed signals and constrained their existence near the GC with PSD$_{ET}$ of 10$^7$ W/Hz among half a million stars (assuming transmitter G$_{ET}\sim10^7$).

\section{Observations}
\label{sect:observations}
\begin{figure*}
    \centering    \includegraphics{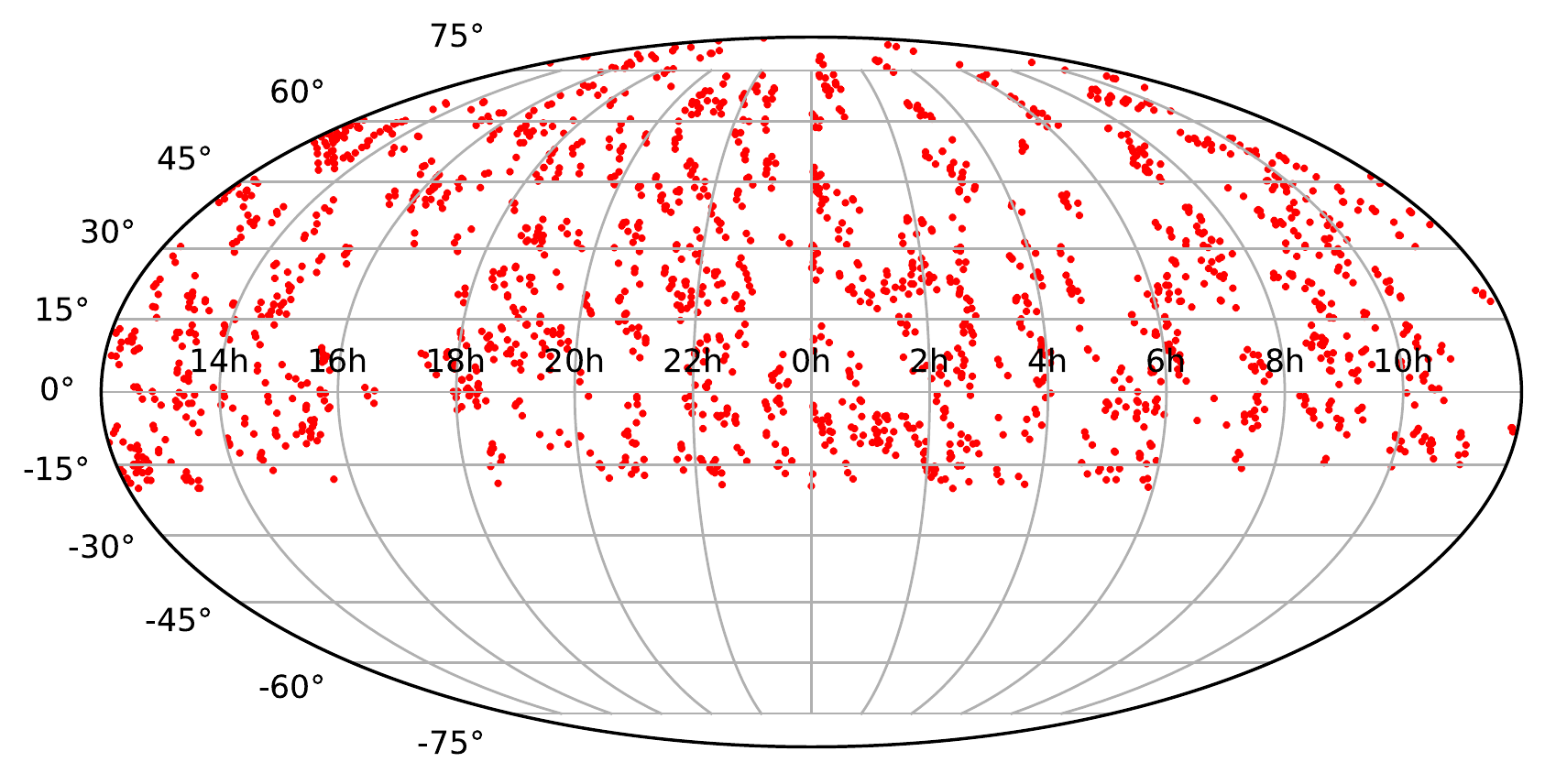}
    \caption{Sky distribution of 1883 observed target stars with the Breakthrough Listen program at the Robert C. Byrd Green Bank telescope across 4--8\,GHz. These sources include a total of 1883 target stars; which includes 595  targets from primary target database and 1288 secondary off-source targets.}
    \label{fig:mollweide_view}
\end{figure*}

We performed our signal search on 233 hours of BL observations conducted between 2017 July to 2018 June at the Green Bank Telescope (GBT) from 4--8\,GHz. These observations collectively provide the largest sample of SETI observations at these frequencies, which are higher than those chosen for the majority of prior searches. The observations employ a position-switching RFI mitigation method, observing a cadence of three ``on-target" scans interspersed with three ``off-target" scans. The ``on" targets are drawn from the BL primary target database which consists of around 1200 nearby stars (see \citealt{ism+17} for details on target selection) for the GBT. In order to improve the efficiency of our observations, the ``off" targets are selected from a secondary list of nearby stars not included in the primary catalog. Here, we report observations from 2795 independent observations of 1883 stars; which includes 595 stars from primary target database and 1288 secondary target stars. For our search, we are not bound to use the similar ``on" and ``off" strategy, hence, we treated 1883 targets as independent observations. Table \ref{tab:observations} provides a truncated list of these targets which also shown in Figure \ref{fig:mollweide_view}. 

We observed these targets with the BL Digital Recorder \citep{MacMahon_2018_BLDR}, which is a state-of-the-art, 64-node, GPU-equipped compute cluster at the GBT. The cluster is divided into 8 banks, with each bank hosting 8 compute nodes. Each compute node records a 187.5\,MHz segment of incoming bandwidth, with each bank recording 1500\,MHz of intermediate-frequency (IF) bandwidth. Our observations used 4 banks with overlapping frequency coverage, providing a total observing bandwidth of $\sim 4875$\,MHz (3563--8438\,MHz) which covers the C-band receiver band of 3950--8000\,MHz. We initially record data as raw baseband voltages in the GUPPI raw format \citep{leb19_bl_data_format}, and then convert these baseband data to total intensity SIGPROC-formatted filterbank files with 364\,kHz spectral and 349\,$\mu$sec temporal resolution. For this study, we further binned these filterbank datasets from 13312 to 6656 frequency channels before searching for aDM beacons.  

\begin{table}[]
    \centering
    \begin{tabular}{lcccccc}
     \hline
     \hline
     Name & RA (J2000) & DEC (J200) & Spectral Type & Distance (pc) & R$_{max}^{50}$ & PSD$_{ET}^{aDM}$ (W/Hz) \\
     \hline
    GJ699 & 17.963222 & 4.739167 & M3.5V & 1.83 & 1701.0 & 9.23e-01 \\ 
    GJ820A & 21.116528 & 38.763889 & K5.0V & 3.49 & 501.0 & 3.36e+00 \\ 
    GJ820B & 21.116919 & 38.755833 & K7.0V & 3.49 & 501.0 & 3.36e+00 \\ 
    GJ280 & 7.654806 & 5.220278 & F5IV & 3.50 & 501.0 & 3.38e+00 \\ 
    GJ15A & 0.307528 & 44.024722 & M1.5V & 3.57 & 501.0 & 3.51e+00 \\ 
    \multicolumn{6}{l}{...} \\
    HIP107727 & 21.823058 & 34.064861 & F8 & 840.34 & 501.0 & 1.95e+05 \\ 
    HIP15159 & 3.256378 & 47.278278 & A4V & 869.57 & 501.0 & 2.08e+05 \\ 
    HIP24467 & 5.251625 & 4.905222 & B8 & 892.86 & 501.0 & 2.20e+05 \\ 
    HIP100753 & 20.427672 & 43.967583 & B8 & 900.90 & 501.0 & 2.24e+05 \\ 
    HIP96852 & 19.686911 & 12.062444 & B0Ib:n & 980.39 & 901.0 & 2.65e+05 \\ 
    \hline
    \end{tabular}
    \caption{The truncated list of targets analyzed in this work's search for broadband pulsed ETI beacons. The columns, from left to right, are: the name of the target star, its right ascension (RA) and declination (DEC) in J2000 coordinates, its stellar spectral type, its distance from Earth, the measured repetitiveness with $\geq$50\% probability, and the estimated putative transmitter output power. The final two columns are defined in Sections \ref{sect:power_budget} and \ref{sect:repetitiveness_of_broadband_beacon}.}
    \label{tab:observations}
\end{table}

\section{\texttt{SPANDAK} pipeline}
\label{sect:SPANDAK}
\begin{figure}
    \centering
    \includegraphics[scale=0.22]{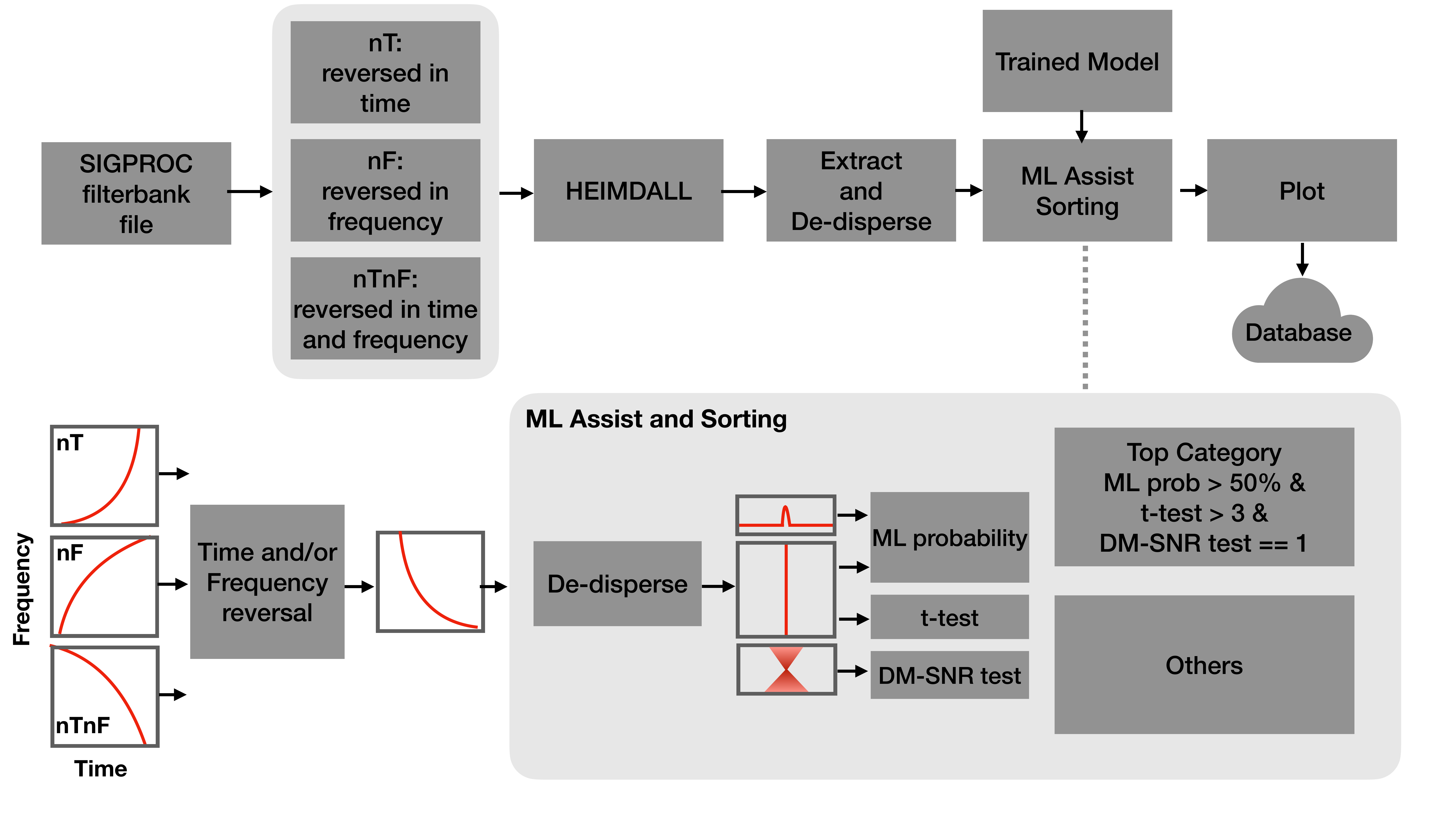}
    \caption{A schematic view of the broadband pulse detection survey pipeline \texttt{SPANDAK}. The input SIGPROC-filterbank files were time or/and frequency reversed to search for the corresponding class of aDM signals. The pipeline uses \texttt{HEIMDALL} \citep{bbb+12} which is a GPU-accelerated tool to search for dispersed pulses. The lightly-shaded grey block in the bottom-right of the diagram represents the newly-developed ML-classifier that prioritizes candidate selection. The criteria used to select top candidates are listed in the darker gray block at the top-right of the ML region.}
    \label{fig:pipeline}
\end{figure}

We developed several tools to search for the three different types of aDM signals shown in Figure \ref{fig:negDM}. The tools comprise an entire pipeline which we refer to as \texttt{SPANDAK}, as shown in Figure \ref{fig:pipeline}. To search for transient signals exhibiting nT-aDM type dispersion, the pipeline reversed the order of received samples to counter [$\mathcal{V}$,T$^T$]. Similarly, \texttt{SPANDAK} reversed the order of frequency channels to search for nF-aDM signals ([$\mathcal{V}^T$,T]). For nTnF-aDM signals ([$\mathcal{V}^T$,T$^T$]), the order of both time samples and frequency channels were reversed. These reversals allow any embedded aDM signal to be detected as a natural pDM signal, enabling use of the large suite of publicly-available tools built to search for single pulses of astrophysical origin. For each of the observed 5-minute long SIGPROC filterbank \footnote{www.sigproc.sourceforge.net} files, we produced three reversed filterbank files corresponding to the intended nT, nF, and nTnF signals searches. We used a GPU-accelerated tool --- named \texttt{HEIMDALL} \citep{bbb+12} --- as the main kernel to search for \textit{dispersed} signals in these reversed filterbank files. We searched all three sets of files in parallel across three NVIDIA GTX Titan XP GPUs to expedite processing. 

For each of the order-reversed files, the pipeline ran \texttt{HEIMDALL} across a DM range of 10 to 5000 pc/cm$^{3}$ with the DM steps selected such that the maximum signal-to-noise (S/N) loss due to incorrect DM was always under 15\%. In principle, it is possible to search for higher aDM signals; however, due to our temporal resolution of 0.3\,ms, inter-channel dispersion smearing greatly reduces our sensitivity for larger aDMs. We searched across a range of pulse widths from 0.3--42\,ms. The pipeline accumulates all transient candidates reported with \texttt{HEIMDALL} from a single order-reversed input file and cross-references various candidate parameters (proximity of arrival times, DMs, S/Ns, widths, etc.). We removed candidates which appear across a large range of DMs within a short interval, allowing us to remove a significant number of false positives due to RFI. A short list of selected candidates were extracted from each corresponding filterbank file for further validation. We time-scrunched the extracted data such that the detected pulse would fit within 2--4 time bins. We also frequency-scrunched the pulse to 16--512 frequency channels based on the detected S/N. An example output plot is shown in the inset in Figure \ref{fig:pipeline} (see Figure 9 of \cite{Gajjar_21_BLGCI} for details). 

For candidate validation, the pipeline produces dedispersed dynamic spectra, where an ideal broadband transient pulse should show up across all observed frequencies. The pipeline then selects an on-pulse window based on the width and arrival time reported from \texttt{HEIMDALL} and extracted on-pulse and off-pulse spectra. We flagged as RFI all channels which were four times the standard deviation in the off-pulse spectra. From the remaining channels, we compared the on-pulse and off-pulse spectral energy distribution using a \textit{t-test}. For a true broadband pulse, the \textit{t-test} should show a significant difference between these spectra. Moreover, a true broadband dispersed signal should show both a peak at the correct DM and a gradual decline in the S/N around nearby DMs. \cite{cm03} outlined a relation where a candidate with a width of W$_{ms}$ at the frequency of $\nu_{GHz}$ across a band of $\Delta\nu_{MHz}$ shows 50\% decline in the S/N across $\Delta{DM}$ with respect to the S/N at a true DM. This approximation can be expressed as:
\begin{equation}
    \Delta{DM}~\approx~506\frac{W_{ms}\nu_{GHz}^3}{\Delta{\nu_{MHz}}}. 
\end{equation}
We dedispersed each of the extracted candidates across a DM range of 3$\times\Delta$DM (with 48 DM steps) to compare the S/N variations to those likely to exist for a true broadband signal. These DM-vs-time plots of S/N were also produced for each candidate for visual inspection. 

Across all observations, we found 133,393 candidates which were initially detected with the \texttt{SPANDAK} pipeline. Visually inspecting this many candidates to identify a potential ETI signal is daunting, and would require significant personnel investment. Thus, as mentioned in Section \ref{sect:AI_to_look_for_ETI}, we have developed a fully-automated CNN-based classifier to vet all detected candidates. This classifier returns probabilities for each candidate being a real broadband transient signal, as opposed to spurious interference. This ML classifier is one of the main components of the \texttt{SPANDAK} pipeline and analysis presented here. {Full details about the ML-assisted candidate prioritization are presented in Appendix \ref{sect:ML_pipeline}.}

\section{Results}
\label{sect:results}

\begin{figure}
    \centering
    \subfloat{
    \includegraphics[scale=0.6]{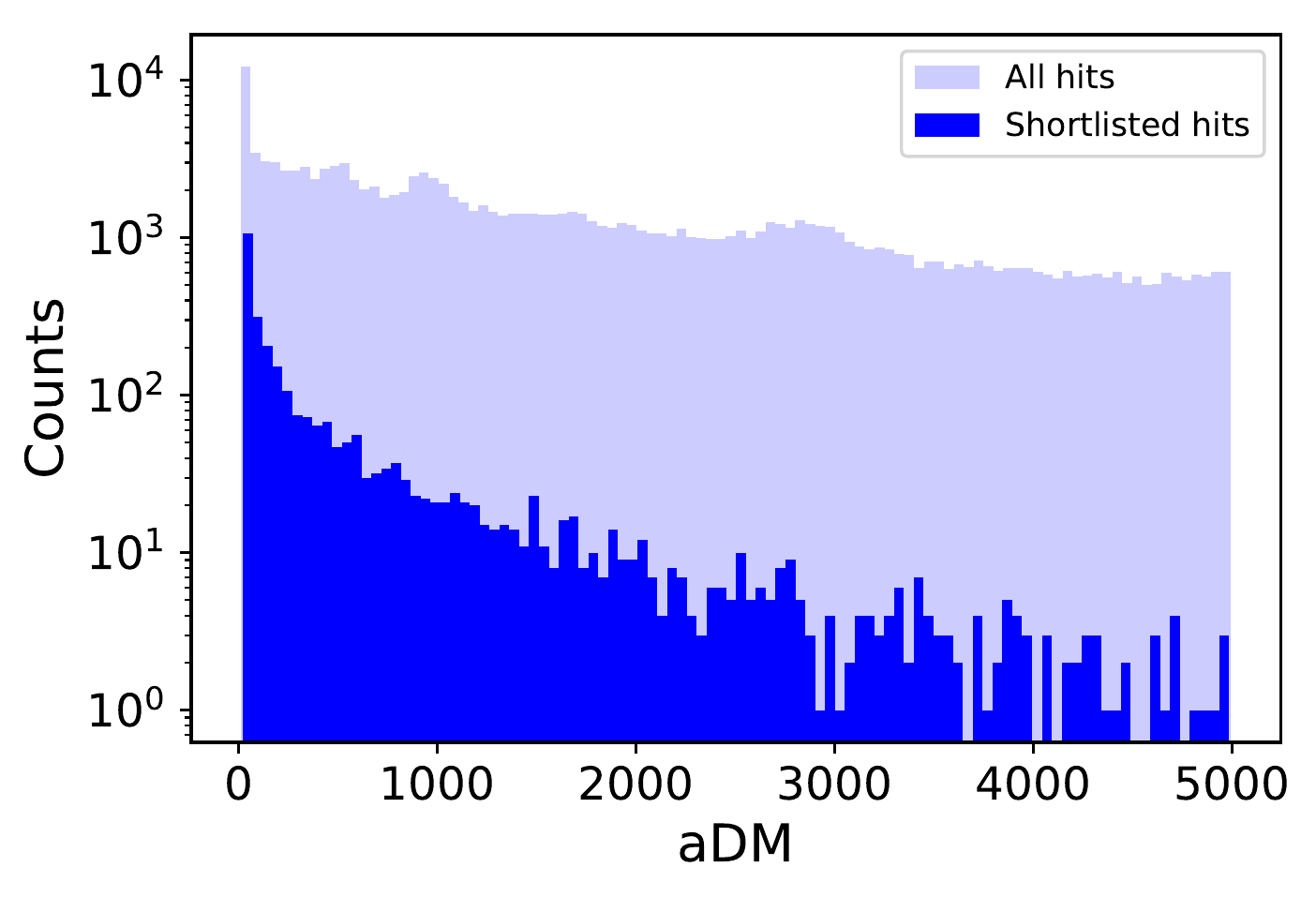}}
    \subfloat{
    \includegraphics[scale=0.6]{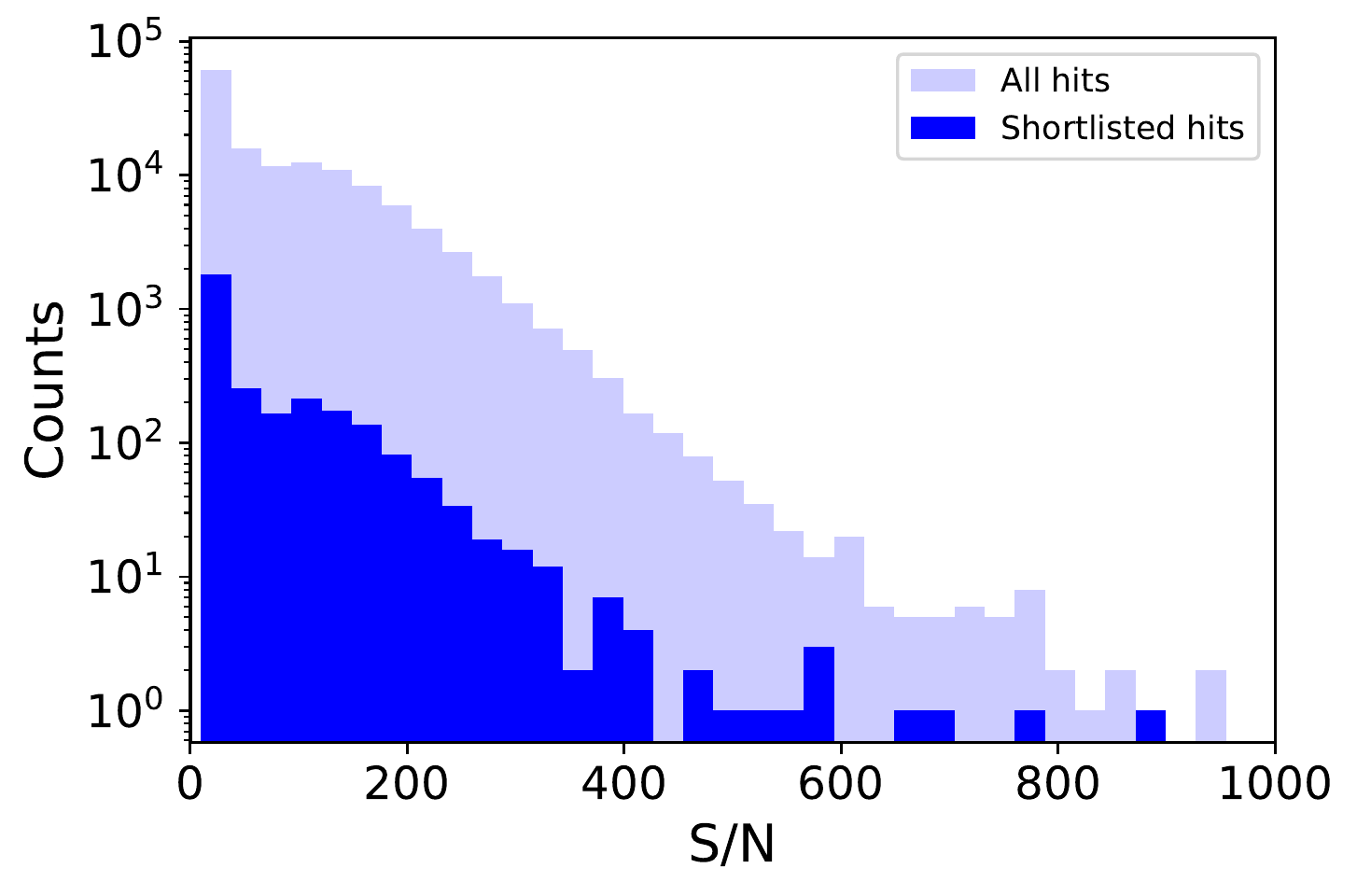}}
    \label{fig:raw_hits_hist} 
    \caption{Distribution of hits detected from the search of broadband pulsed aDM beacons towards 1883 stars. (a) Histogram of measured aDM. A sharp peak near zero DM indicates a large number of these candidates originating due to RFI. (b) Histogram of measured S/N with the \texttt{SPANDAK} pipeline for all detected hits. For both histograms, the lighter blue color shows raw hits, while the darker blue color shows shortlisted final hits.} 
\end{figure} 

We searched $\sim$233 hours of total observations which were divided into 2795 scans, each 5 minutes long. This corresponds to approximately 1883 unique stars (see Table \ref{tab:observations}) comprising 595 primary targets likely to be observed more than once and 1288 secondary stars. Our search for three different classes of aDM signals, using the \texttt{SPANDAK} pipeline, found 133,393 raw hits. Figure \ref{fig:raw_hits_hist} shows the distribution of these hits as a function of aDM and S/N detected with the \texttt{SPANDAK} pipeline. We received an overabundance of hits near zero DM, which is an indication that a large number of the hits are due to local interference. 

We further shortlisted these candidates using various selection criteria mentioned in Section \ref{sect:SPANDAK}. We only selected candidates which showed a Student {\itshape  t-test} value larger than 3.0, analogous to seeing a difference in the on-pulse and off-pulse energy distributions with a {\itshape p-value} $\geq$ 0.99. We then considered ML probabilities utilizing the CNN model described in {Appendix \ref{sect:ML_pipeline}.} We only selected candidates for which our ML probabilities were larger than 50\%, given that the ML model reported an accuracy of around 98\%. This helped us significantly reduce the number of likely false-positives, resulting in 2948 final hits which were visually inspected. Distributions of these shortlisted hits are also shown in Figure \ref{fig:raw_hits_hist}. The ML-assisted shortlisting resulted in a significant reduction in the number of hits across all aDMs. 

\begin{figure}
    \centering
    \subfloat{
    \includegraphics[scale=0.23]{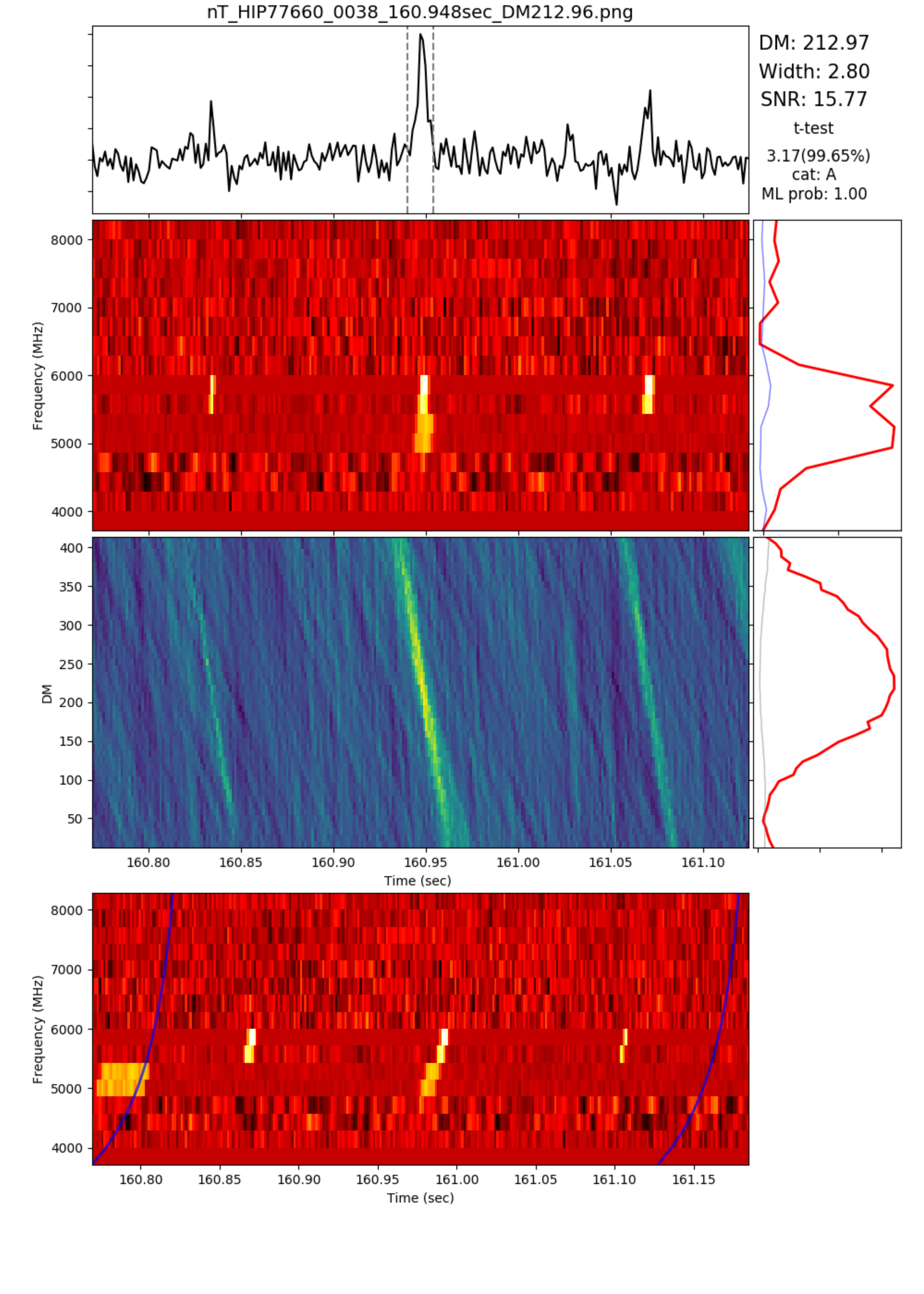}
    \label{fig:example_cand1}}
    \subfloat{
    \includegraphics[scale=0.23]{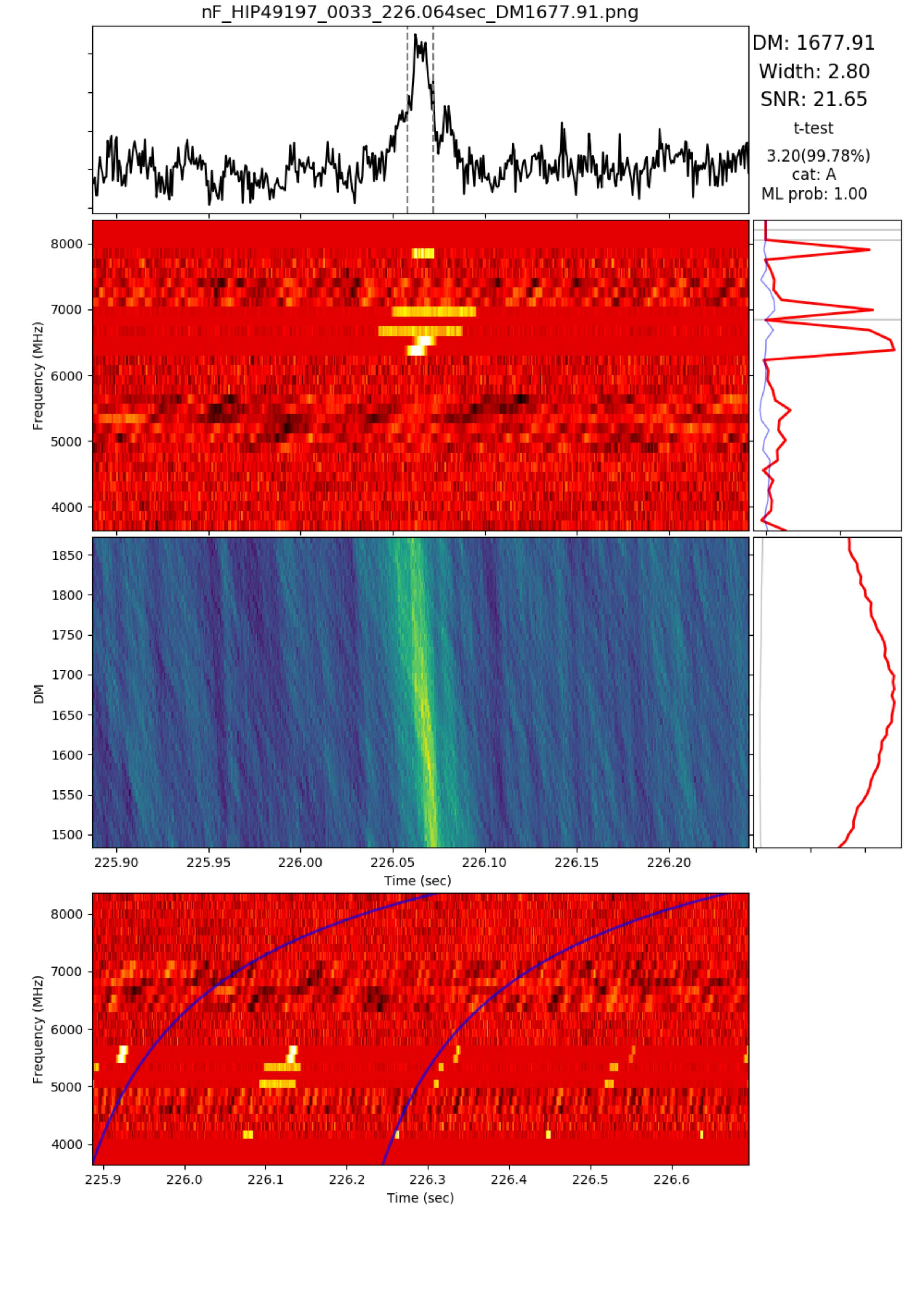}
    \label{fig:example_cand2}}
    \subfloat{
    \includegraphics[scale=0.23]{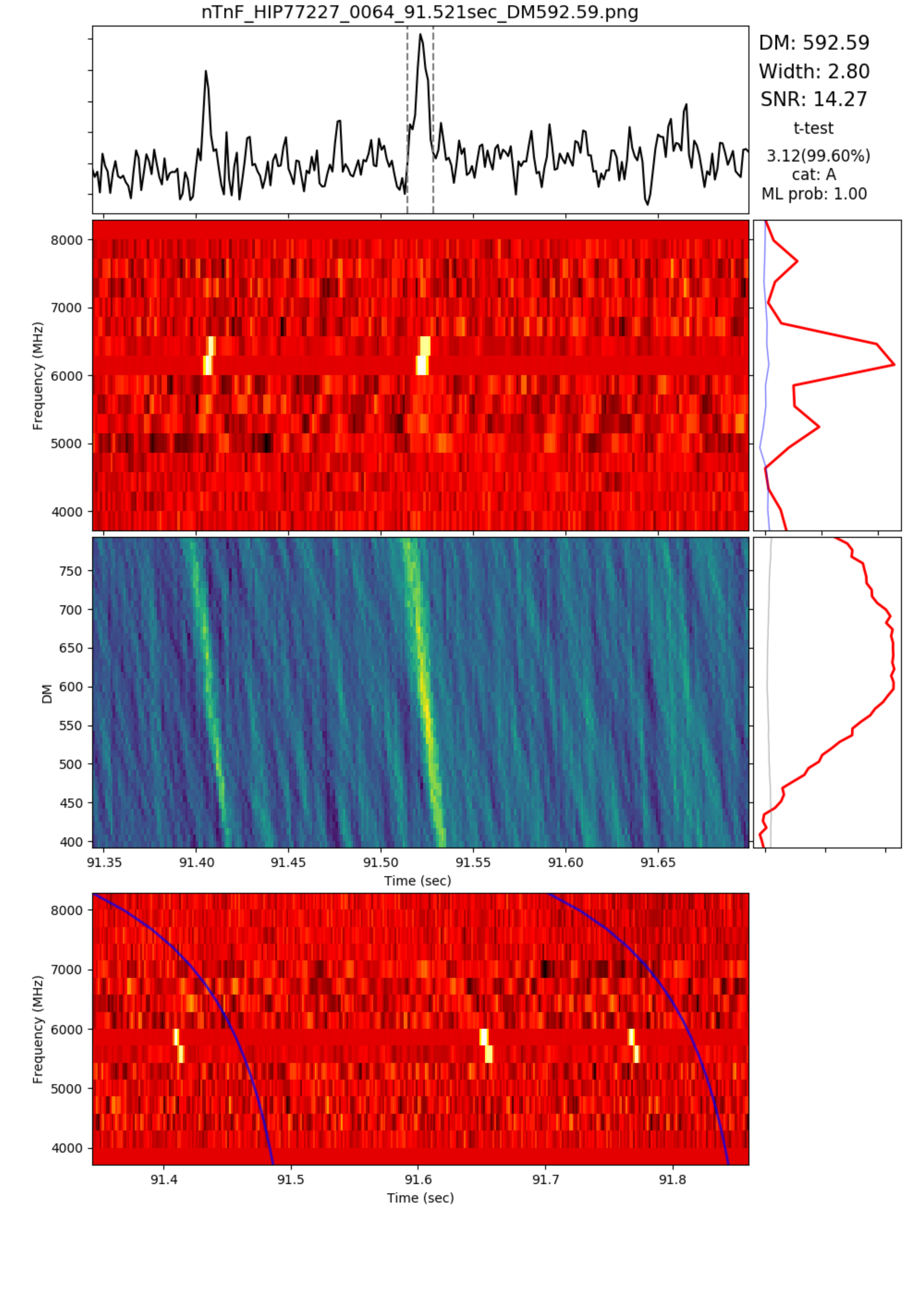}
    \label{fig:example_cand3}}
    \caption{The top candidates in the three different aDM categories found in our search for broadband ETI beacons towards 1883 stars. For each candidate plot, the top panel shows a dedispersed single pulse with an on-pulse region marked by two dotted lines along with the necessary diagnostic information to its right. The second panel from the top shows dedispersed dynamic spectra containing a broadband pulse across 4--8 GHz, with on-pulse and off-pulse spectra shown on its right in red and gray lines, respectively. The third panel from the top shows the DM-vs-time plot, with on-pulse DM-vs-SNR shown in the adjacent panel on the right. The bottom panel shows the original data that triggered the hit, containing the detected dispersed pulse and two blue lines indicating the best-fit DM for visual guidance. The color in all three panels shows normalized intensities in arbitrary units.
    }
    \label{fig:all_example_cands}
\end{figure}

Figure \ref{fig:all_example_cands} shows examples of our top candidates in each of the three aDM classes. Each of these candidates appears to show a dispersion relation loosely matching various aDM beacon criteria. However, we have found similar signals across a large number of targets. This indicates that such candidates arise due to coincidental alignment of spurious temporal interference, which is apparent once the data are compared with the best-fit dispersion curves in the bottom panel of Figure \ref{fig:all_example_cands}. It is likely that the fully automated time and frequency binning in the candidate plots might not be optimum. We have also developed a special interactive tool\footnote{\url{https://github.com/stevecroft/bl-interns/tree/master/jianic}} by which we can iterate over a combination of frequency and time bins for any given candidate plot to improve S/N to better aid with its identification. We visually vetted all 2948 candidates but did not find any signals of interest which we could not rule out as spurious interference and thus, no signals required further inspection through our interactive tool. 
Through our survey, we are therefore able to place probabilistic limits on the presence of ETI beacons towards 1883 stars. 

\section{Discussion}
\label{sect:discussion}
\subsection{Survey Sensitivity}
\label{sect:discussion_survey_sensitivity}
We can estimate our survey's sensitivity for all our targets using Equation \ref{eq:P_et_broad}, solving for the transmitter's power density (PSD$_{ET}$). Figure \ref{fig:sensitivity_and_repeat} shows a histogram of constrained PSD$_{ET}$ for broadband aDM class signals. The histogram shows a bimodal distribution, reflecting the fact that our survey included two sets of targets: primary targets which were within 50\,pc from Earth and secondary targets which extended all the way up to 1000\,pc. The median PSD$_{ET}$ from all observed targets is around 10$^3$ W/Hz, but the lowest expected PSD$_{ET}$ is on the order of 1 W/Hz for our closest source, GJ\,699. By comparison, powerful aircraft radar\footnote{\url{www.mobileradar.org/radar_descptn_3.html}}, which also emit powerful broadband pulses (200 MHz wide), have power densities on the order of 10$^{-3}$ W/Hz.

\subsection{Repetitiveness of Broadband Beacons}
\label{sect:repetitiveness_of_broadband_beacon}

\begin{figure}[!ht]
    \centering
    \subfloat{
    \includegraphics[scale=0.45]{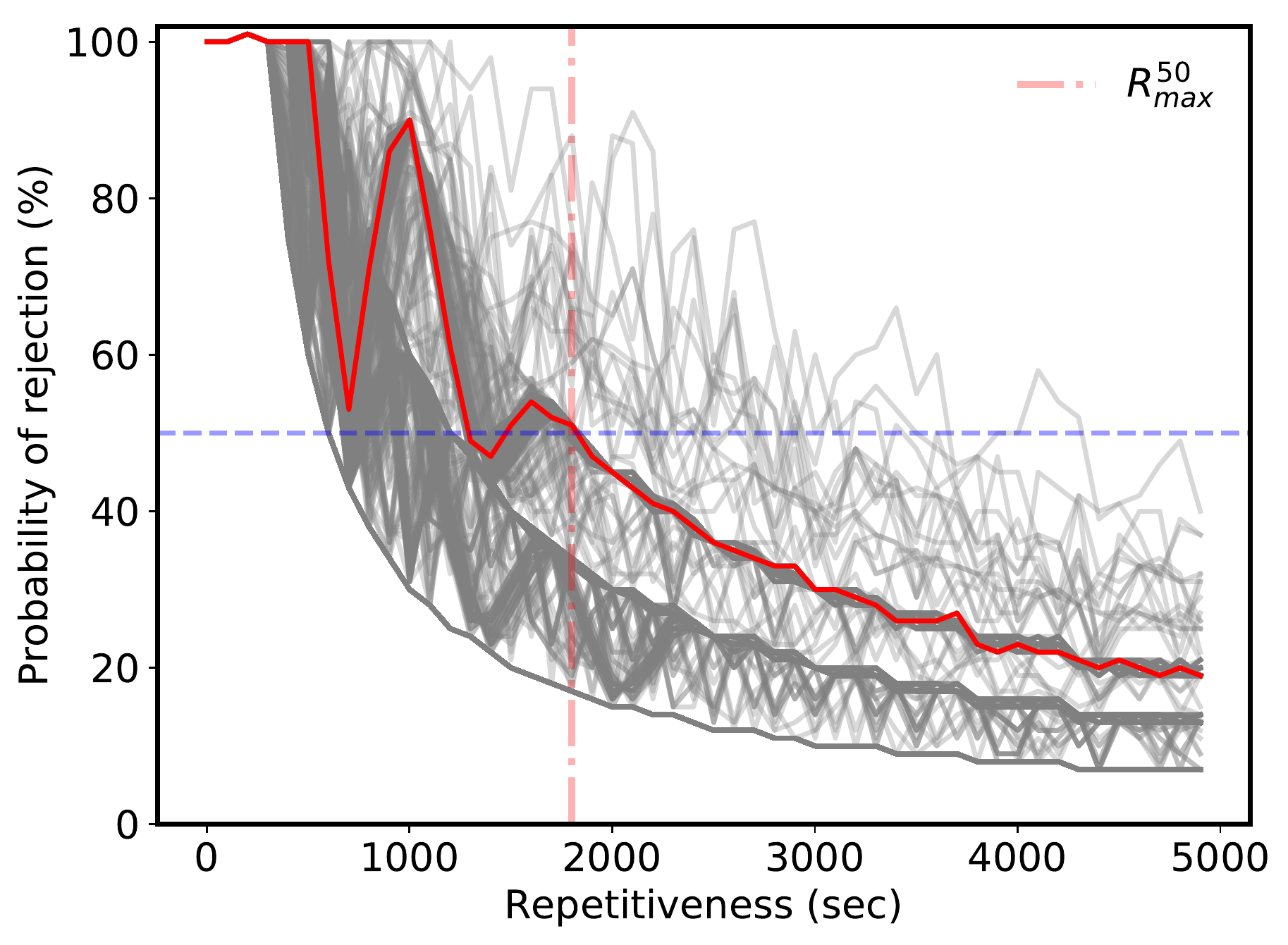}
    \label{fig:repetitiveness}}
    \subfloat{
    \includegraphics[scale=0.45]{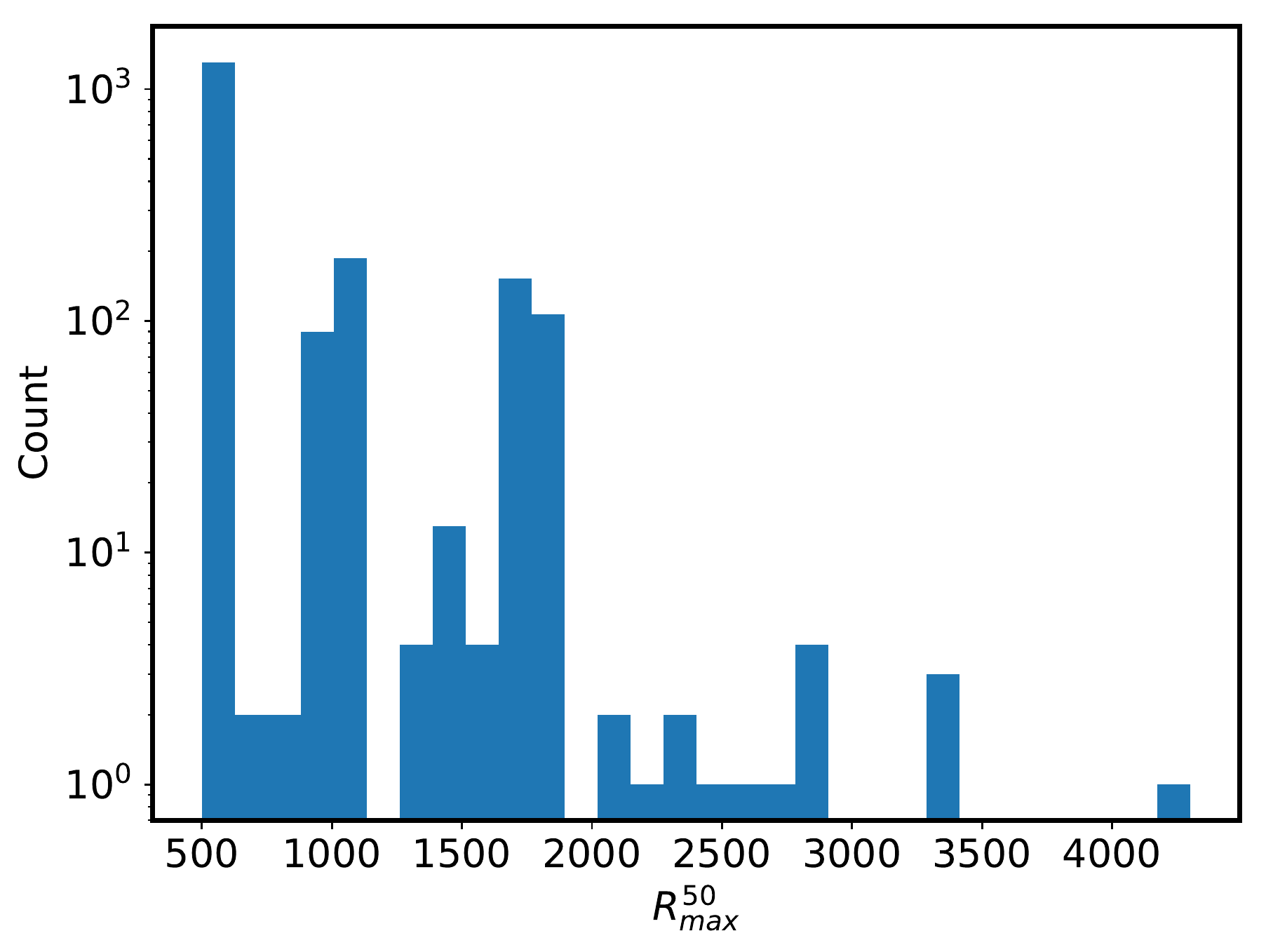}
    \label{fig:rmax}}
    \caption{Repetitiveness distribution from the survey of broadband ETI beacons. {\itshape Left:} Probability of rejection for a range of repetitiveness of broadband beacons for 1883 targets shown with gray solid lines. The red solid line shows a rejection probability for a typical set of three on-source and three interspaced off-source observations, each 5 minutes in length. The dashed blue and dashed-dotted red lines show 50\% rejection probability at the corresponding R$_{max}$, respectively. {\itshape Right:} Histogram of measured R$_{max}^{50}$ from all 1883 targets. Peaks around 500, 1100, and 1800 seconds in the measured R$_{max}^{50}$ are apparent corresponding to sources observed for a total of 5, 10, and $\geq$15 minutes, respectively. A small number of targets were observed more than three times, populating the tail of the histogram.}
    \label{fig:my_label}
\end{figure}

As mentioned in Section \ref{sect:power_budget}, broadband pulsed beacons are expected to repeat pulses in order to increase chances of detection. Through our observations, we can constrain the repetitiveness of these broadband aDM signals. For example, from a single 5-minute long observation, we can reject with high probability the notion that the repetitiveness of aDM broadband signals is under 300 seconds. For many of targets included in this analysis, we are likely to have observed them for multiple 5-minute scans; interspaced by 5-minute scans of different targets. To measure the probability of rejection for such a set of observations, we simulated and arranged broadband pulses on the time axis and then simulated a pulse train with a range of repetition periods. 

We also adjusted the phase (or offset) of these bursts within the corresponding period under consideration. These pulse trains of different offsets were overlapped with observations for a given target on the same time axis. We counted the number of instances which would have allowed us to detect at least one pulse for a given repetition period across all offsets. For periods where we were able to detect at least one pulse for all phase offsets, we can reject such repetitions with near 100\% accuracy. For periods where we were able to detect one pulse for half of the offsets, we can only reject repetitions with 50\% accuracy. Figure \ref{fig:repetitiveness} shows the probability of rejection of repetitiveness from this above-mentioned exercise across all observed sources.

We measure R$_{max}^{50}$, as shown in Figure \ref{fig:repetitiveness}, as the maximum period for which repetitiveness can be rejected with more than 50\% probability for each target. Figure \ref{fig:rmax} shows a histogram of measured R$_{max}^{50}$ for all targets. As mentioned in Section \ref{sect:observations}, our observations towards primary targets were conducted with multiple 5-minute on-source scans. As shown in Figure \ref{fig:repetitiveness}, for observations of primary targets with three 5-minute long on-source scans, the R$_{max}^{50}$ is approximately 1800 seconds. For most of the secondary targets, which were observed only once, a R$_{max}^{50}$ peak exists at around 500 seconds. There were also some stars (primary and secondary) which were observed two times, which leads to an R$_{max}^{50}$ peak at around 1100 seconds. A few primary targets were observed more than three times on different days, which leads to a constraint on longer repetitiveness (see Figure \ref{fig:repetitiveness}) and extended tail in the R$_{max}^{50}$ histogram shown in Figure \ref{fig:rmax}. 

\subsection{Broadband transmitter occurrence rate and Drake equation}
\label{sect:ETI_transmitter_rate}
\begin{figure}
    \centering
    \includegraphics[scale=0.7]{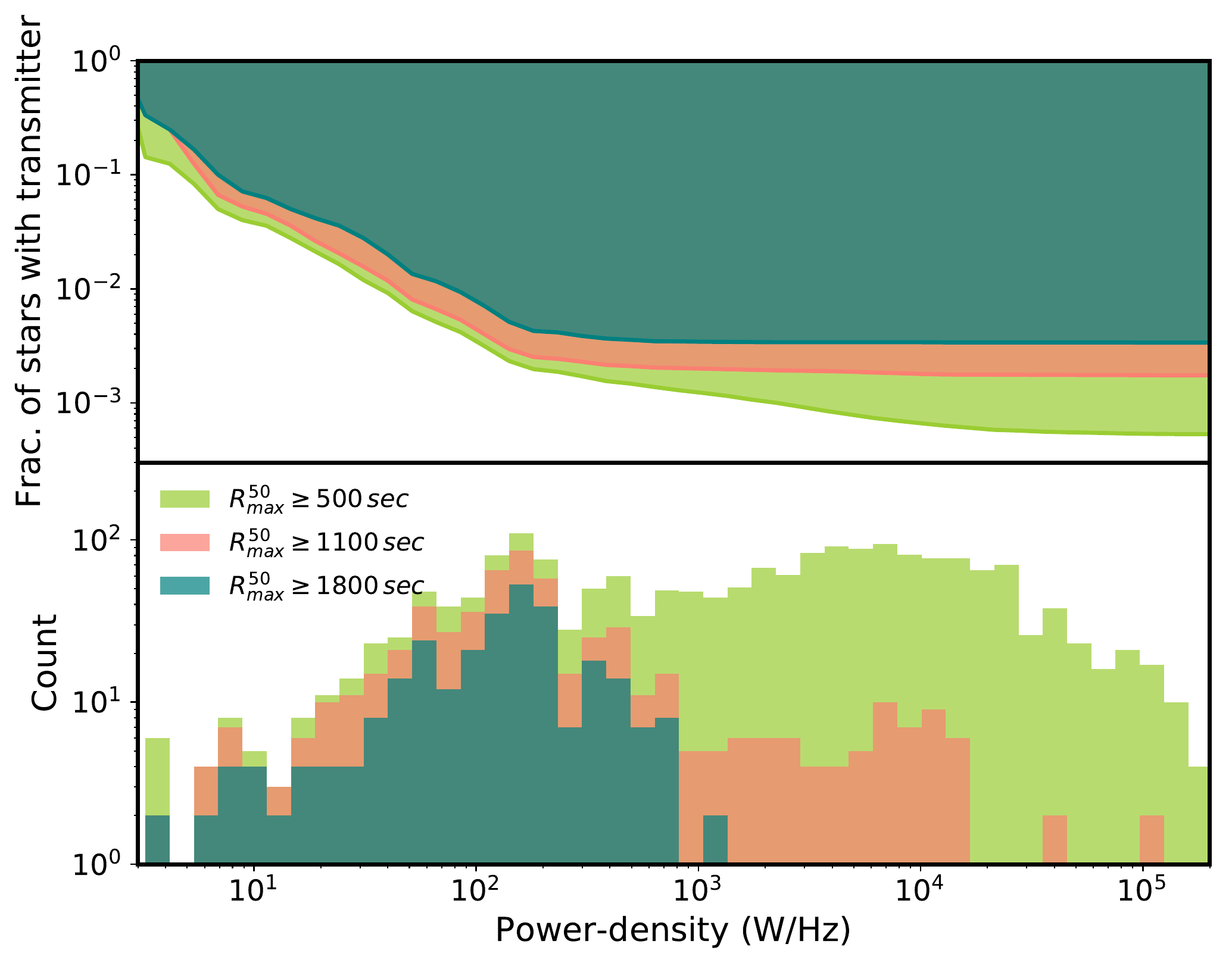}
    \caption{Transmitter rate and power density histograms for broadband pulsed ETI beacons of all types. The bottom panel shows a histogram of power density limits obtained towards all our targets, subdivided into three subgroups of R$_{max}^{50}$ (i.e. repetitiveness with $\geq$50\% probability). The top panel shows cumulative distributions of these subgroups, shown with the shaded region, representing the fraction of stars hosting a transmitter across a range of power densities. With our survey, we can reject an occurrence rate of more than 1 in 1000 stars in the solar neighborhood transmitting a broadband ETI beacon with PSD$_{ET}\gtrsim$ 10$^5$ W/Hz with a repetition rate of less than 500 seconds.}
    \label{fig:sensitivity_and_repeat}
\end{figure}

In recent years, radio SETI has been quickly expanding the search, eliminating regions of parameter space which are unlikely to host ETI beacons. \cite{Enriquez:2017} introduced  a metric known as the transmitter rate for narrowband ETI beacons, which represents the number of transmitters per star as a function of EIRP (see Figure 7 in \citealt{Enriquez:2017}). \cite{Gajjar_21_BLGCI} carried out a search for broadband ETI beacons in a blind survey towards the Galactic Center and placed a limit of $\lesssim$1 in approximately half a million stars with PSD$_{ET}\gtrsim$ 10$^7$ W/Hz. It should be noted that the average required PSD$_{ET}$ from that study was several orders-of-magnitude larger than this survey (see Section \ref{sect:discussion_survey_sensitivity}). Here, we have placed some of the very first constraints on the fraction of stars with broadband ETI beacons in the solar neighborhood across PSD$_{ET}$ ranging from 1 to $10^5$ W/Hz. Figure \ref{fig:sensitivity_and_repeat} shows histograms of constrained PSD$_{ET}$ (i.e., transmitted power densities) subgrouping them into different R$_{max}^{50}$ across all targets. We then measured the cumulative distribution of these counts, represented by the shaded areas in Figure \ref{fig:sensitivity_and_repeat}. This figure shows the region of the underlying parameter space our survey was able to reject for repeating broadband ETI beacons. For example, we can reject an occurrence rate of more than 1 in 1000 stars in the solar neighborhood transmitting a broadband ETI beacon with PSD$_{ET}\gtrsim$ 10$^5$ W/Hz with a repetition rate of less than 500 seconds. 

\begin{figure}
    \centering
    \includegraphics[scale=0.7]{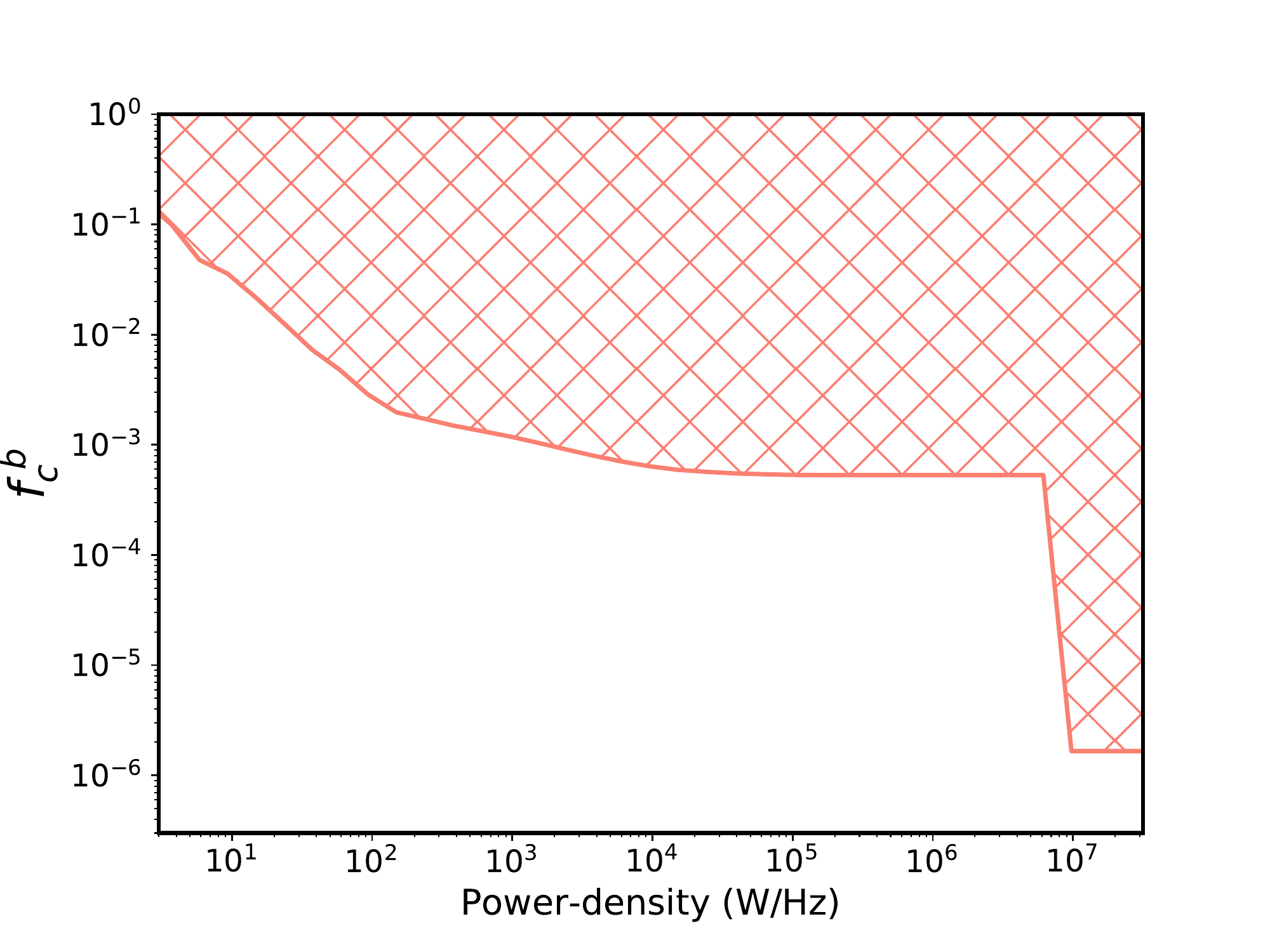}
    \caption{Fraction of stars producing broadband signals with artificial dispersion ($f_c^b$) as a function of PSD for the entire Milky Way. The hashed region in red represents constraints from 1883  nearby stars from our current survey and around half a million stars located at the Galactic Center from a previous survey by \cite{Gajjar_21_BLGCI}.}
    \label{fig:fcb_combined}
\end{figure}

The Drake Equation \citep{shklovskii_sagan_book} provides a simple metric to estimate a speculative abundance of communicative ETI civilizations in the Milky Way we are likely to detect. For broadband signals with artificial dispersion, this estimation (N) can be approximated as 
\begin{equation}
\rm{N~=~R_{IP}\;f_{c}^b\;L}
\label{eq:drake}
\end{equation}
Here,  $\rm R_{IP}$ is the emergence rate ($\rm yr^{-1}$) of technologically advanced intelligent life in the Milky Way; $\rm f_{c}^b$ represents the fraction of these advanced civilizations producing broadband signals with artificial dispersion; and $\rm L$ is the lifetime of such a civilization. The $\rm R_{IP}$ is a combination of the average rate of star formation and the fraction of stars providing suitable conditions and time for life to emerge and evolve. This rate is similar but not strictly related to the formation rate of planets inside the conventional habitable zones around different spectral-type stars. Moreover, due to the differential distribution of metallicity and the history of star formation in the Milky Way, $\rm R_{IP}$ can be speculated to be widely different for different parts of the Milky Way. For example, \cite{lineweaver2004GHZ} suggested that there exists a Galactic Habitable Zone (GHZ), which is an annulus extending from 7 to 9\,kpc from the Galactic Center providing an ideal location for advanced life to emerge and evolve. Thus, the solar neighbourhood and more broadly the spiral arms of the Milky Way are expected to have a relatively high $\rm R_{IP}$. Contrary to the \cite{lineweaver2004GHZ} model, \cite{Morrison:2015} and \cite{Gajjar_21_BLGCI} argued that the Galactic Center is expected to have higher $\rm R_{IP}$ due to the sheer number of stars. Although it is hard to estimate $\rm R_{IP}$ and L, conducting surveys like ours can help us to jointly constrain them from the inferred limits on the fraction of stars producing signals of our interest; i.e. $\rm R_{IP}\times{L}\leq{1/f_{c}^b}$. Combining constraints from our current survey with those from \cite{Gajjar_21_BLGCI}, which searched for similar classes of signals at the center of the Milky Way, enables us to estimate $\rm f_{c}^b$. In our current survey, which extends up to 1\,kpc from the Sun, we can roughly constrain $\rm f_{c}^b$ in the solar neighbourhood, or more generally, in the GHZ speculated by \cite{lineweaver2004GHZ}. Similarly, from \cite{Gajjar_21_BLGCI} we can get similar constraints on $\rm f_{c}^b$ at the center of the Milky Way. Figure \ref{fig:fcb_combined} shows a combined constraint $\rm f_{c}^b$ from these two surveys, which can provide one of the most stringent limits on the fraction of stars producing broadband beacons in the spiral arms and center of the Milky Way.

\subsection{Future work}

In the analysis presented here, we searched for dispersed signals which followed a frequency and time relation corresponding with a dispersion index of 2.0. It is plausible that a true signal from ETI might exhibit a dispersion index that is not exactly 2.0. We are continuing to explore that parameter space by searching for artificially dispersed signals exhibiting other dispersion indices using an expansion of the ML techniques presented in this paper. Moreover, in this work, we performed a search for broadband beacons reliant on bright individual pulses (i.e., with a flat period prior). \cite{Sullivan_91_ETI_pulse_periods} gave a list of potentially unique periods for broadband pulsed beacons which could indicate artificiality, including the lifetime of a neutron ($\sim$896 seconds) or the lifetime of the most luminous optical line OII (46.7 seconds). Thus, in the future, we plan to carry out a full periodicity search for these aDM signals. This will also allow us to detect even weaker broadband signals, as we can fold the underlying time series to improve S/N.

\vspace{5cm}
\section{Conclusion}
\label{sect:conclusion}
Radio SETI has so far been largely focused on searches for narrowband CW signals. We demonstrate that broadband pulsed beacons are energetically efficient compared to CW signals given longer operational timescales. We carry out one of the first comprehensive surveys for this newly-suggested class of ETI beacon towards 1883 stars by searching for broadband pulsed beacons with artificial or negative dispersion. This search used 233 hours of data taken across 4--8\,GHz with the Robert C. Byrd Green Bank telescope. We used a GPU-accelerated pipeline named \texttt{SPANDAK} to search for three different classes of broadband signals; nT--aDM, nF--aDM, and nTnF--aDM. We found $\sim$10$^5$ initial hits from our filter-based search approach. To reduce the number of false-positives, we locally designed and deployed a fully-automated CNN-based classifier. We trained this classifier to identify dedispersed broadband beacons, and used it to prioritize the hits from all three classes of dedispersed aDM beacons. To the best of our knowledge, this is one of the first uses of an ML-based approach for radio SETI across such a large number of targets. With the assistance of the ML classifier we were able to reduce the number of false-positives by 97\%. We did not detect an aDM beacon of artificial nature in our datasets. Hence, we place a constraint on the existence of broadband pulsed beacons in our solar neighbourhood with $\lesssim$1 in 1000 stars exhibiting a PSD$_{ET}\gtrsim$ 10$^5$ W/Hz and repeating $\leq$500 seconds. 

\section{ACKNOWLEDGMENTS}
Breakthrough Listen is managed by the Breakthrough Initiatives, sponsored by the Breakthrough Prize Foundation. We thank the staff at the Green Bank observatory for their operational support. We would also like to thank the anonymous referee for all their suggestions which helped us improve our draft. S.Z.S. acknowledges that this material is based upon work supported by the National Science Foundation MPS-Ascend Postdoctoral Research Fellowship under Grant No. 2138147.
\bibliographystyle{aasjournal}
\bibliography{references_new,references,mybib_frb}

\appendix 
\renewcommand\thefigure{\thesection.\arabic{figure}}
\section{Machine Learning-assisted candidate prioritization}
\setcounter{figure}{0}
\label{sect:ML_pipeline}

As illustrated in Figure \ref{fig:pipeline}, we build a single CNN classifier\footnote{https://github.com/DominicL3/hey-aliens} to characterize all three types of aDM signals. We achieve this by training the ML classifier to classify dedispersed dynamic spectra and frequency-averaged pulses, rather than building three different classifiers for three different classes of aDM signals. Moreover, the same classifier can also be used to search for FRBs, which will be discussed in future publications (Gajjar et al. 2022 in prep). This concept is partially similar to that presented by \cite{Agarwal_2020_FETCH}. In the following subsections, we outline our simulation to train the classifier, model architecture, and recovery tests. 

\subsection{Simulating dedispersed broadband signals}
Modern supervised learning methods require large amounts of training data, so we created a synthetic dataset of a wide variety of mock broadband signals embedded in the background of real telescope noise and spurious RFI. We randomly selected 40 observations (each 5 minutes long) from our set of 2795 observations. This randomly selected sample of observations is likely to represent the necessary backgrounds for most of our observations. 
To simulate a training set of broadband transient signals, we followed a similar procedure to \cite{Connor_2018} with a few modifications. We first randomly selected start times from one of the 40 observations and extracted the appropriate number of samples corresponding to a randomly selected dispersion delay (DM). Before injecting a simulated broadband signal on top of a real observation, we dedispersed the empty background to this randomly selected DM. This allows our network to see the telescope background after ``dedispersion", since it is not at zero DM and thus more likely to represent the backgrounds of real dedispersed signals. This dedispersed background was time-scrunched to 256 bins and frequency-scrunched to 16 channels, providing a good compromise between data resolution and memory requirements during training. 

\begin{figure}[h]
    \centering
    \includegraphics[width=0.9\textwidth]{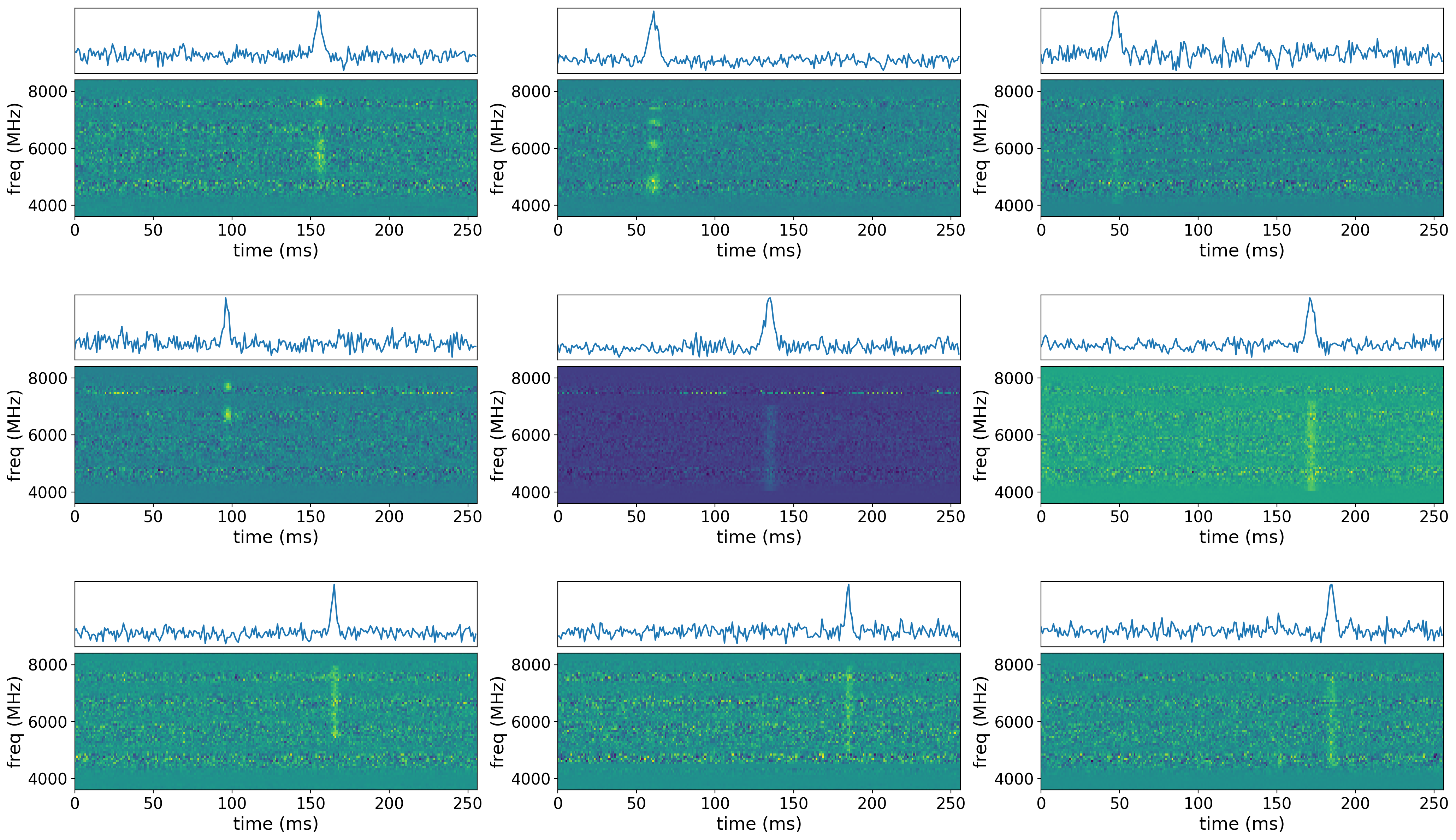}
    \caption{Six examples of dedispersed simulated broadband ETI beacons injected into real observations. In each plot, the top panel shows a dedispersed time-series containing the pulse while the bottom panel shows the dedispersed dynamic spectrum. We simulated $\sim$100,000 such examples, using real observations as the background, in order to train our ML-classifier.}
    \label{fig:simulated_FRBs}
\end{figure}

For our simulation, we first produced a broadband dedispersed signal of constant intensity across all observed channels, akin to a frequency-averaged Gaussian pulse. The S/N of this signal was sampled from a log-normal distribution ranging from 6 to 20. The spectrum of the simulated pulse was convolved with a cosine function of random phase to mimic any frequency-dependent intensity variations. To simulate the effect of frequency channels being flagged and removed due to excessive RFI, we randomly removed 10\% -- 50\% of channels. The resulting broadband pulse was then added to the "dedispersed" background. Finally, each signal and background is also over-dispersed or under-dispersed by a small percentage, sampled from a normally distribution with a standard deviation of 0.005; this helped us imitate inaccuracies in our search pipeline that could report slightly wrong pDM values. A total of around 200,000 data points were simulated, in which 100,000 of these synthetic data contained mock broadband signals embedded in real telescope backgrounds, and the remaining 100,000 contained empty backgrounds. Figure \ref{fig:simulated_FRBs} visualizes nine randomly selected broadband pulses, showing the dedispersed waterfall plot along with its frequency-averaged time series. We shuffled our simulated dataset and used 50\% for training and 50\% for validation. We discuss the network architecture in the following sections. 

\subsection{Model Architecture} 
\label{sect:Model_architecture}
\begin{figure}
    \centering
    \includegraphics[scale=0.3]{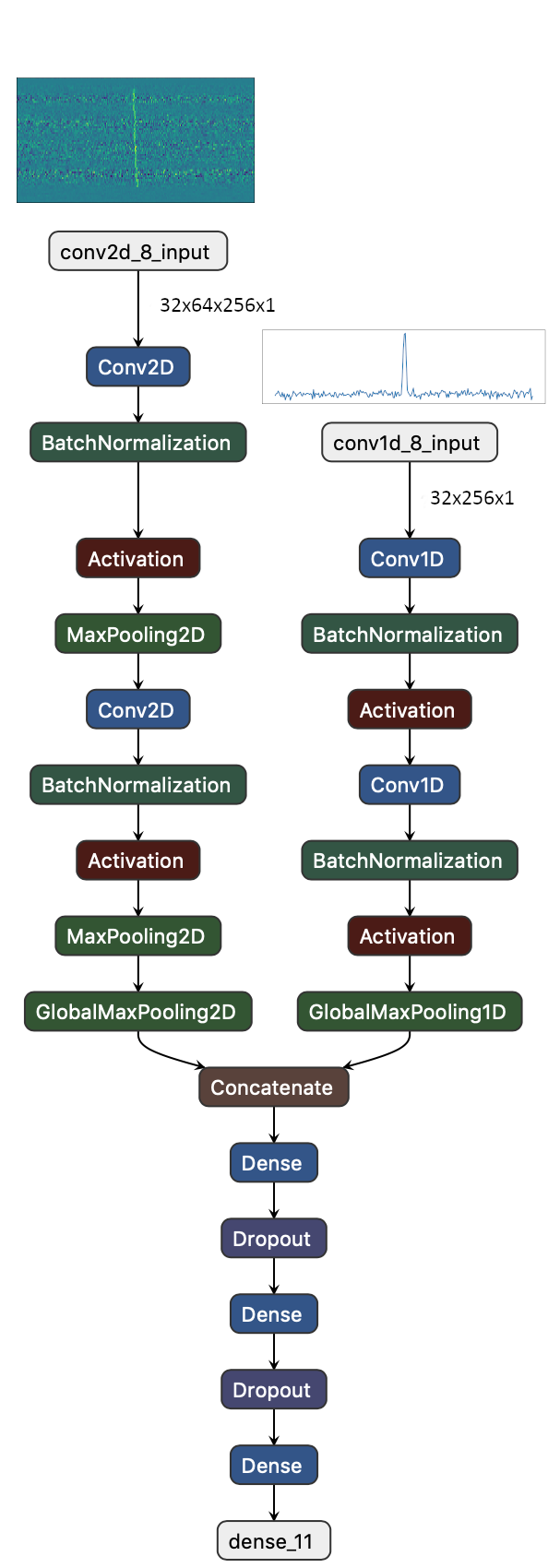}
    \caption{The neural network model architecture of the ML-classifier built to identify dedispersed broadband ETI beacons. The model was trained on two different characteristics of the broadband pulse, corresponding to the two branches: the left branch is the spectrogram branch, which takes an input of a two-dimensional dynamic spectrum, while the right branch takes an input of a one-dimensional time-series. Both branches were concatenated to assess final probability (see Section \ref{sect:Model_architecture} for details).}
    \label{fig:model_architecture}
\end{figure}

Compared to state-of-the-art CNNs today, our model is very simple, as the task at hand is not complex and can be likened to detecting a vertical/near vertical line in an image. For every example, the model takes two inputs, a 2D dynamic spectrum, and its corresponding 1D frequency-averaged time series (similar to examples shown in Figure \ref{fig:simulated_FRBs}). In Figure \ref{fig:pipeline}, these inputs are shown in the middle and top inset plots, respectively. These inputs are fed into two separate branches of the network --- what we call the spectrogram branch and the time series branch --- which extract features from the inputs and are eventually concatenated together to produce one softmax prediction. Figure \ref{fig:model_architecture} contains a visualization of the network architecture, with the spectrogram and time series branches on the left and right, respectively.

We found that an architecture with 2 convolutional layers worked best for our application, retaining the ability to recognize signals without being unnecessarily complex. A 3x3 kernel with a stride length of 1 ensures that the convolution operation does not ``skip over" broken broadband signals. In the time series branch, the number of filters in a convolutional layer is roughly half the number of filters in the parallel convolutional layer in the spectrogram branch to prevent overfitting, since detecting whether a peak is present in a 1D signal is not a hugely complex task. Each convolutional layer is followed by a \texttt{BatchNormalization} layer and an \texttt{ReLU} activation. We use \texttt{MaxPooling} layers at the end of every convolutional block in the spectrogram branch to reduce the dimensionality, but found that \texttt{MaxPooling} layers within the time series branch led to losing the peaks in the 1D signal, and thereby decided to remove them. After all convolutional blocks for both branches, we use global max pooling to transform all convolutional feature maps into a 1D tensor for each example. Subsequently, the two branches of the network are fused by concatenating the outputs of the \texttt{GlobalMaxPooling} layers, after which they are fed into two fully-connected layers before the prediction layer. We also use \texttt{Dropout} layers between the fully-connected layers, which have been shown to be a simple method of reducing overfitting \citep{10.5555/2627435.2670313}.

\subsection{Training parameters}
\begin{figure}[h]
    \centering
    \includegraphics[width=0.7\linewidth]{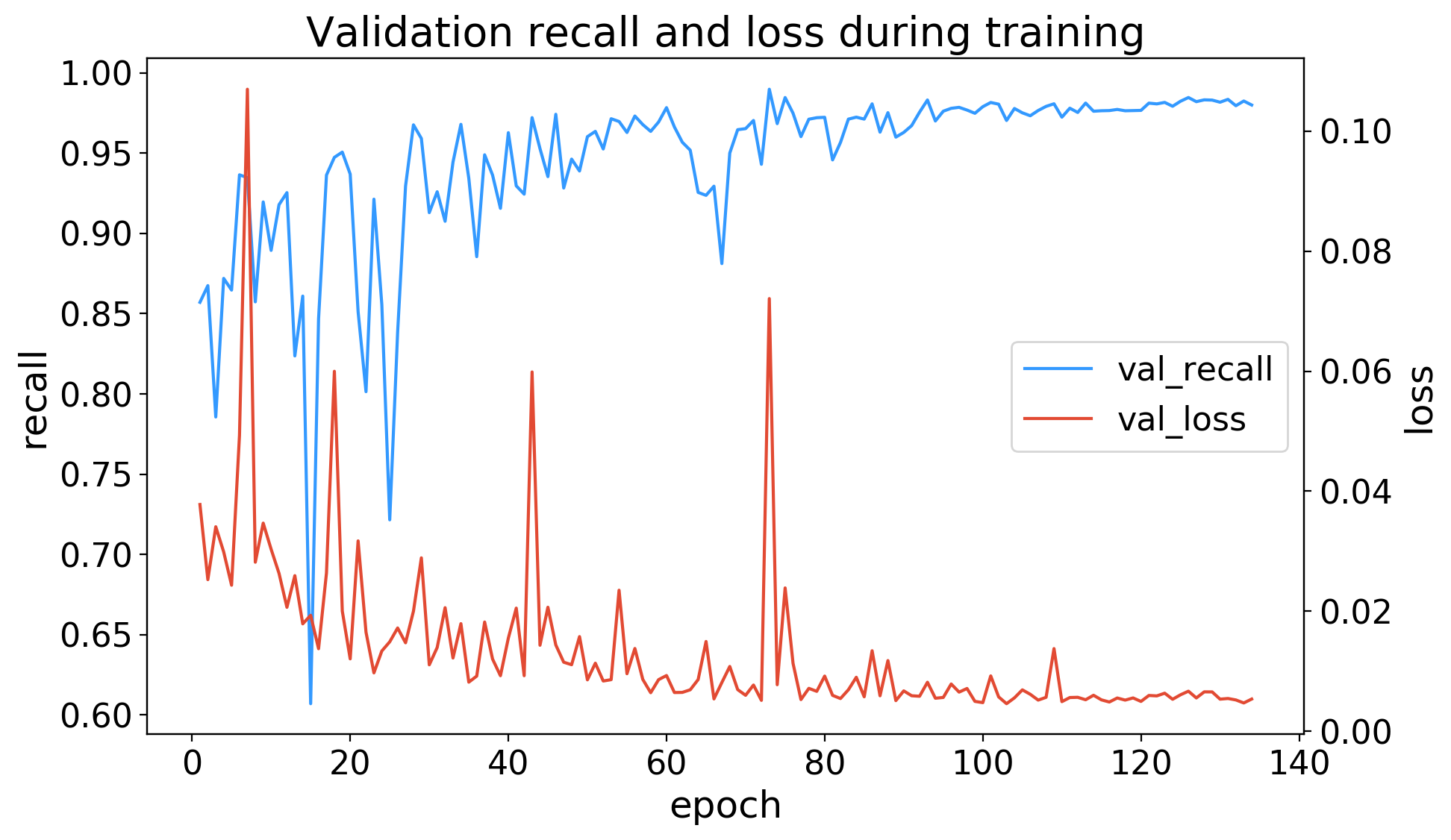}
    \caption{Recall and loss for the validation set over 134 epochs from the broadband transient detection ML pipeline. The left abscissa represents the recall rate from the simulated datasets, while the right abscissa represents the corresponding loss function. The ordinate represents epoch number, which appears to produce constant recall and loss beyond 110 epochs. With the help of this supervised ML-classifier, we were able to reject 97\% of false positives.}
    \label{fig:recall_during_training}
\end{figure}

Preprocessing of the data is done on a per-array basis. For each array, we subtract the median from each row (the spectrum) and divide the entire array by its standard deviation. As stated above, data were split evenly between training and validation sets, such that 100,000 data points went to training and the other 100,000 went to validation.

We implement our classifier in Keras 2.0.8 and TensorFlow 1.4.1, compiling the model with a binary cross-entropy loss and the Adam optimizer \citep{kingma2017adam}. Because we value recall over precision, we weight the positive class 10 times as much as the negative class, thereby penalizing the network more for missing signals than for false positives.

We employ several Keras callbacks to supplement model training:
\begin{itemize}
    \item \texttt{ModelCheckpoint}: save the model only when the validation loss decreases from its last known minimum, thus only saving the model that produces the lowest validation loss.
    \item \texttt{ReduceLROnPlateau}: halve the learning rate if validation loss doesn't improve after 15 epochs.
    \item \texttt{EarlyStopping}: stop training if validation loss does not improve after 30 epochs.
\end{itemize}

With a batch size of 32, we trained our models using a single Nvidia Titan XP GPU. Though we allowed the model to run for a maximum of 500 epochs, \texttt{EarlyStopping} halted training after 134 epochs after not seeing a decrease in validation loss in the designated number of epochs. For the model presented in this paper, training completed the 134 epochs in about 5.5 hours. Figure \ref{fig:recall_during_training} displays the recall and loss curves for the validation set over the course of training 134 epochs. Our model converges with a validation recall of 0.9897. 



\end{document}